\documentclass[twocolumn]{aastex631}

\usepackage{xspace}
\usepackage{subfigure}
\usepackage{xcolor}
\usepackage{xr}
\newcommand{\wow}{Wow!\@ Signal\xspace}

\newcommand*{\HI}{\textsc{Hi}\xspace}
\newcommand*{\HII}{\textsc{Hii}\xspace}

\begin{document}
\raggedbottom

\title{Arecibo Wow! I: An Astrophysical Explanation for the \wow}
%\version{1}
\author[0000-0002-0786-7307]{Abel M\'endez}
\affiliation{Planetary Habitability Laboratory, University of Puerto Rico at Arecibo}%, PO Box 4010, Arecibo, PR 00613, USA}

\correspondingauthor{Abel M\'endez}
\email{abel.mendez@upr.edu}

\author[0000-0003-3455-8814]{Kevin Ortiz Ceballos}
\affiliation{Center for Astrophysics ${\rm \mid}$ Harvard {\rm \&} Smithsonian}%, 60 Garden St, Cambridge, MA 02138, USA}

\author[0000-0002-6140-3116]{Jorge I. Zuluaga}
\affiliation{SEAP/FACom, Instituto de F\'{\i}sica - FCEN, University of Antioquia}%Calle 70 No. 52-21, Medell\'in, Colombia

\begin{abstract}
The Ohio State University Big Ear radio telescope detected in 1977 the \wow, one of the most famous and intriguing signals of extraterrestrial origin. Characterized by its strong relative intensity and narrow bandwidth near the 1420 MHz hydrogen line, its source has never been detected again despite numerous follow-up attempts. 
Arecibo Wow! is a new technosignature project using archived data from the Arecibo Observatory. Here we present our first results of drift scans made between February and May 2020 at 1420 MHz. The methods, frequency, and bandwidth of these observations are similar to those used to detect the \wow. However, our observations are more sensitive, have better temporal resolution, and include polarization measurements. 
We report the detection of narrowband signals ($\Delta\nu\leq$ 10 kHz) near the hydrogen line similar to the \wow, although two-orders of magnitude less intense and in multiple locations. Despite the similarities, these signals are easily identifiable as small interstellar clouds of cold hydrogen (\HI) in the galaxy. We hypothesize that the \wow was caused by a sudden brightening of the hydrogen line in these clouds triggered by a strong transient radiation source, such as a magnetar flare or a soft gamma repeater (SGR). A maser flare or superradiance mechanisms can produce stimulated emission consistent with the \wow.
Our hypothesis explains all observed properties of the \wow, proposes a new source of false positives in technosignature searches, and suggests that the \wow could be the first recorded event of an astronomical maser-like flare in the hydrogen line.
\end{abstract}

% https://astrothesaurus.org/concept-select/
\keywords{Search for extraterrestrial intelligence (2127), Neutral hydrogen clouds (1099), Interstellar masers (846), Magnetars (992), Radio astronomy (1338), History of astronomy (1868)}

\section{Introduction \label{sec:intro}} 

The \wow, detected on August 15, 1977, by the Ohio State University's Big Ear radio telescope, remains one of the most compelling potential instances of extraterrestrial communication \citep{shuch_wow_2011}. The signal, identified by astronomer Jerry R. Ehman, exhibited a narrowband radio frequency of approximately 1420.456 MHz, which is notably close to the hydrogen line frequency, a region of the radio spectrum often considered a natural candidate for interstellar communication \citep{Kraus1979}.

The signal lasted for 72 seconds, corresponding to the time it took the telescope beam to sweep past the source, but no follow-up observations have successfully replicated this anomaly \citep{Gray2002}. The transient nature of the \wow, combined with its strength and frequency, has led to extensive debate and investigation within the SETI community (Search for Extraterrestrial Intelligence), yet its exact origin remains undetermined (see, e.g., \citealt{Tarter2001}). 

Further analysis of the \wow suggests that its strength and narrow bandwidth are inconsistent with natural astronomical sources, such as pulsars or known interstellar phenomena, thereby raising the possibility of an artificial origin \citep{Shostak2003}. The \wow remains an enigmatic outlier, representing a tantalizing hint of potential extraterrestrial communication while underscoring the challenges inherent in distinguishing genuine signals from cosmic noise and terrestrial interference \citep{Tarter2010, kipping2022could}.

Despite numerous efforts using various radio telescopes, including the Very Large Array (VLA), no repeat detections have been made at the original frequency or any other candidate frequencies \citep{gray_vla_2001,Backus2004, perez2022breakthrough}. The search for similar signals continues to be a significant focus within the SETI community, employing advanced digital signal processing techniques and expanded observational capabilities to improve the chances of detecting faint and transient signals \citep{Werthimer2001}.

The Allen Telescope Array (ATA) was also used to search for possible repetitions \citep{2020AJ....160..162H}. The researchers monitored the specific frequency and sky coordinates associated with the original signal, using the ATA's capabilities to capture a broad frequency range in search of potential extraterrestrial signals. Despite comprehensive observation efforts spanning multiple sessions, the study did not result in the detection of any repeat signals. This outcome suggests that the \wow may have been a unique occurrence or an artifact of terrestrial interference, rather than a signal of extraterrestrial origin.

The \wow is frequently described as narrowband in contrast to the broadband spectra typical of most astronomical radiation. However, it is crucial to note that the \wow bandwidth is around 10 kHz, which is ten times the bandwidth of some astrophysical masers \citep{rajabi2020astronomical} and a thousand times greater than the bandwidth of signals targeted in numerous contemporary radio SETI searches \citep{2022AcAau.190...24W,2022AcAau.199..166H}. For example, UCLA SETI handles data with a 3 Hz resolution \citep{2023AJ....166..206M}.

The main objective of this paper is to present the first results of our Arecibo Wow! project. We include a comparison between the Ohio State University (Ohio SETI) and the Arecibo Observatory SETI projects, which should provide a context of the \wow and our observations (\autoref{sec:wow} and \autoref{sec:seti-arecibo}). Arecibo Wow! is a side project of Arecibo REDS, and they are both described in \autoref{sec:awow}. We present the analysis of our observations in \autoref{sec:analysis}. Our proposed mechanisms for the origin of the \wow is described in \autoref{sec:mechanism}. Finally, we discuss the implications of this study in \autoref{sec:discussion} and summarize our study findings in \autoref{sec:conclusion}.

\section{Ohio SETI\label{sec:wow}}

The Big Ear radio telescope, located at Ohio State University, played a pivotal role in the Ohio Sky Survey prior to 1973 \citep{1970AJ.....75..351E}. This project was dedicated to measuring the positions and strengths of wide-band radio sources operating primarily within the 21-cm hydrogen line at a bandwidth of 8 MHz, spanning from 1411 to 1419 MHz. During the course of the survey, approximately 20,000 radio sources were cataloged, nearly half of which were previously unobserved by any other radio telescope. Subsequent observations were made at other observatories to refine the positions of these sources, and several were optically identified, including two that were recognized as the most distant quasars known at the time \citep{Kraus1979}.

In 1973, following a shift in funding priorities by the United States Congress, the National Science Foundation (NSF) withdrew its financial support for the Ohio Sky Survey, which had a significant impact on the project. The loss of funding led to the cessation of the Ohio Sky Survey and the eventual disbandment of the survey team. Despite these setbacks, the Big Ear radio telescope was repurposed for a new mission: the systematic search for narrowband radio signals, or Ohio SETI.

Unlike wide-band radio sources, which are typically natural and generate emissions across a broad spectrum (including radio, optical, X-ray, and gamma-ray bands), narrowband sources are often artificial, such as those used in AM, FM, TV broadcasts, satellite communications, and radar, implying potential intelligent origin. With the narrowband search underway, a systematic survey of the declination range from -36° to +64° began \citep{1977Icar...30..267D}.

\begin{figure*}
    \centering
    \includegraphics[width=1\linewidth]{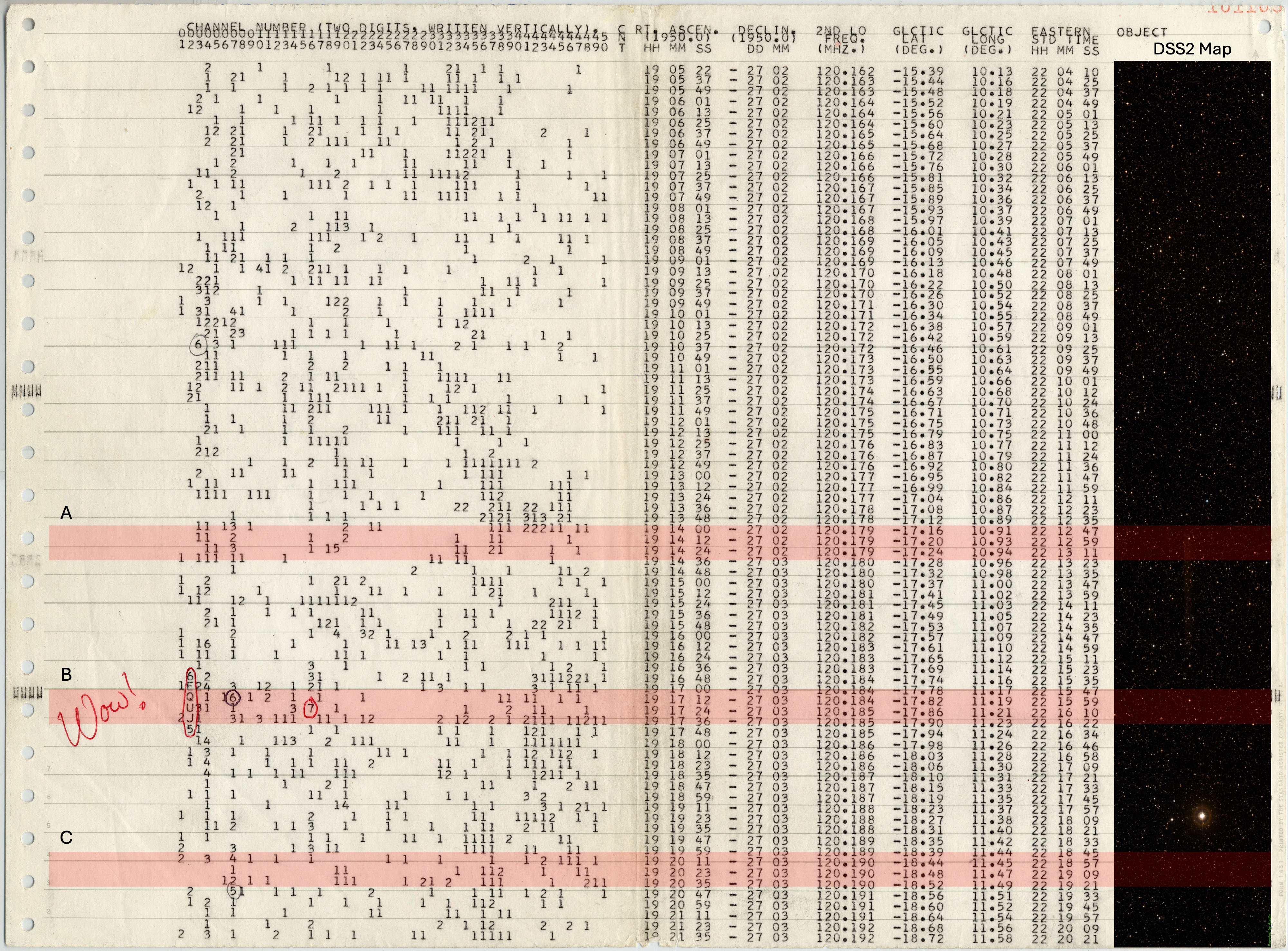}
    \caption{Full page of the computer printout with the \wow labeled in handwritten red ink. The coordinates column corresponds to the location of the positive horn. The signals are calculated from the signal-to-noise ratio (SNR) of the absolute value of the difference between the positive (ON) and negative (OFF) horn, which is 45 arcminutes ahead (3 minutes). Three locations of interest were added to the figure, labeled A, B, and C, and highlighted in red. They have the approximate azimuth size of the Big Ear telescope (8 arcminutes). The actual location of the \wow is either in B or C. If the signal was present and persistent in B, then the signal would appear in both A and B, but there is nothing in A. If present and persistent in C, then it would appear in both B and C, but there is nothing in C. A Digitized Sky Survey 2 (DSS2) frame was included for reference with the objects in the observed field (right). Printout credit: \href{https://ohiomemory.org/digital/collection/p267401coll32/id/12429/}{The Ohio History Connection Collections}. \label{fig:printout}}
\end{figure*}

The data collected by the Big Ear radio telescope were processed and recorded in a series of computer printouts, each row encapsulating approximately 12 seconds of sidereal time (\autoref{fig:printout}). During each 10-second period, the telescope collected one intensity value per second across 50 channels, averaging these ten values to represent the intensity of each channel throughout the 10-second integration period \citep{1977Icar...30..267D}.

The remaining 2 seconds of the 12-second interval were dedicated to computer analysis, where it processed the data and identified any notable phenomena. The printouts displayed the average intensities for each of the 50 channels (each having $\Delta\nu=10$ kHz), arranged sequentially from left to right, with channel 1 on the left side and channel 50 on the right side. 

% Figure: Wow! Signal

Intensity values were converted to a single character for display: zero values were rendered as blanks, integers 1 to 9 were printed directly as digits, and integers 10 to 35 were represented by uppercase letters A through Z, respectively. Each character corresponds to a specific range of signal-to-noise ratios. During a standard analysis of the data gathered on August 15, an unusual character string, ``6EQUJ5,'' was detected in channel 2 of the printout on August 19, 1977.

This sequence was immediately recognized as indicative of a narrowband radio source with a small angular diameter, a potential signal of intelligent origin. The strongest signal, represented by the character ``U," indicates that the signal was 30.5 $\pm$ 0.5 times stronger than the background noise. This signal was later referred to as the \wow. In \autoref{tab:wow} we describe its characteristics, including location, time, and absolute intensity.

% Table: Wow Description

\begin{deluxetable*}{ll}
\tablecaption{Description of the \wow. \label{tab:wow}}
\tablehead{
\colhead{Parameter} & \colhead{Value}
}
\startdata
Date & August 15, 1977 \\
Time & 22:16:01s (10:16:01 PM) EST (02:16 UTC, August 16) \\
Location & Sagittarius Constellation \\
Frequency & 1420.4556 $\pm$ 0.005 MHz \\
Observation Frame & Galactic Center of Rest (GCR) \\
Bandwidth & narrowband ($\le$ 10 kHz) \\
Signal Strength & 30.5 $\pm$ 0.5 times background noise (SNR $\approx$ 30 to 31) \\
Duration & 72 seconds \\
Positive (West) Horn Coordinates (Equatorial) & RA: 19h 25m 31s $\pm$ 10s, Dec: -26d 57m $\pm$ 20m (J2000) \\
Negative (East) Horn Coordinates (Equatorial) & RA: 19h 28m 22s $\pm$ 10s, Dec: -26d 57m $\pm$ 20m (J2000) \\
Positive (West) Horn Coordinates (Galactic) & lon: 11d 39.0m $\pm$ 0.9m, lat: -18d 53.4m $\pm$ 2.1m \\
Negative (East) Horn Coordinates (Galactic) & lon: 11d 54.0m $\pm$ 0.9m, lat: -19d 28.8m $\pm$ 2.1m \\
Estimated Intensity & $\approx$ 54 or 212 Jy \\
\enddata
\tablerefs{\citep{shuch_wow_2011}}
\end{deluxetable*}

The hypothesis that has gained the most attention, especially among the public, is that the \wow was a transmission from an extraterrestrial civilization \citep{wright2021strategies}. The frequency of the signal, close to the hydrogen line, is considered a logical choice for interstellar communication because it is naturally quiet and would be universally recognized by technologically advanced civilizations \citep{drake1973interstellar}. However, this hypothesis faces critical issues.

The signal was never repeated, despite numerous follow-up observations. In the context of SETI, repetition is essential for verification, as a one-time detection makes it difficult to rule out other sources. Moreover, the absence of modulation or encoding in the signal, which would be expected from an intentional transmission, further complicates this interpretation.

Another possibility is that the \wow originated from a natural astrophysical source. Candidates include stellar flares, pulsars, or other cosmic phenomena that emit radio waves (for a recent explanation overview, see \citealt{kipping2022could}). However, these sources typically exhibit characteristics such as periodicity or broader bandwidths, which the \wow lacks.

The narrowband nature of the signal is particularly unusual because natural emissions are generally broadband. Additionally, the hydrogen line frequency is not usually associated with strong (non-thermal) radio emissions from astrophysical objects, making this hypothesis less likely.

% ISS
The possibility of interstellar scintillation (ISS) has also been raised as a potential explanation for the \wow. In the case of ISS, a radio signal can fluctuate significantly in strength and phase when passing through the ionized interstellar medium \citep{cordes_interstellar_1991,cordes_scintillation-induced_1997,brzycki_detecting_2023}. A key obstacle is that ISS will not make a broadband signal narrowband and the probability of a scintillation gain on the order of a hundred is very small \citep{gray_vla_2001}.

Terrestrial interference is another plausible explanation. The signal could have been caused by a reflected terrestrial transmission or an Earth-based signal that temporarily overwhelmed the receiver. The 1420 MHz band is protected for astronomical purposes, but illegal or accidental transmission in this band could have occurred. However, the Big Ear radio telescope had a design that helped filter out terrestrial sources, and the signal appeared only in one of the two feed horns, complicating the explanation of interference. No known terrestrial source has been identified that matches the characteristics of the \wow.\footnote{For a description and some of the hypotheses that have been raised about the terrestrial and extraterrestrial origin of the signal see the detailed account written by Jerry R. Ehman himself in 2010 at \href{http://www.bigear.org/Wow30th/wow30th.htm}{this URL} (last visited: \today).}

One hypothesis suggests that the signal could have been caused by a reflection of Earth-based signals off space debris or satellites \citep{ehman2011wow}. This idea posits that a radio signal from Earth might have bounced off a satellite or space debris and was then picked up by the telescope. However, this explanation struggles with the same issue of non-repetition. 

The duration of the signal and its specific characteristics do not align neatly with the known satellite reflections or the behavior of the debris. Additionally, the Doppler shift expected from such a reflection was not observed. Moreover, a reflecting piece of space debris must have been moving very slowly and not tumbling to mimic a point source in the celestial sphere, which is unlikely \citep{shuch_wow_2011}.

There is also the consideration of interstellar propagation effects, where the signal could have been distorted or focused through gravitational lensing or other space phenomena. These effects could potentially create a signal that mimics the characteristics of the \wow.  However, the gravitational lensing hypothesis suffers from the drawback of a short signal duration. The \wow was not detected in the second horn of the Big Ear whose beam swept the same area of the sky only 3 minutes later \citep{shuch_wow_2011}. Additionally, such a scenario would likely involve complex interactions with multiple cosmic objects, which are improbable and would likely have been observed again under similar conditions. 

Finally, human error has been considered in the recording or processing of the data. Data handling methods in 1977 were not as advanced as today, and it is conceivable that a mistake or anomaly in the data processing chain produced the \wow. However, this explanation lacks specific evidence and does not account for the signal's characteristics, which seem consistent with a real astronomical event.

In summary, any explanation of the \wow must account for its five principal characteristics \citep{ehman2010big}: (1) it exhibited a narrowband nature, with a bandwidth equal or less than 10 kHz, (2) the mean signal strength remained invariant for a minimum of 72 seconds, persisting for a duration ranging from minutes to hours, as it was detected in only one of two sequential horns (separated by a three-minute interval) and not within a 24-hour period succeeding this event, (3) it possessed a high flux density, measured in the range of tens to hundreds of Janskys, (4) no modulation was discerned on a temporal scale shorter than 10 seconds, and (5) it was a singular observation, not observed in subsequent observations by Big Ear or any other telescopes henceforth.

% Table: Radio Telescopes

\begin{deluxetable*}{lcc}
\tablecaption{Comparison of the telescopes of the Arecibo Observatory (AO) and the Ohio State University Radio Observatory (OSURO). \label{tab:telescopes}}
\tablehead{
\colhead{Property} & \colhead{Arecibo Telescope} & \colhead{Ohio State University Telescope (Big Ear)}
}
\startdata
Telescope Type & Fixed dish with moving receiver & Fixed paraboloid with meridian-transit \\
Operational Years & 1963 - 2020 & 1963 - 1998 \\
Current Status & Decommissioned (collapsed in 2020) & Dismantled in 1998\\
Location & Arecibo, Puerto Rico, USA & Delaware, Ohio, USA \\
Longitude & 66.7528° W & 83.0336° W \\
Latitude & 18.3442° N & 40.2567° N \\
Elevation & 497 m & 276 m \\
Main Reflector & Spherical Reflector & Cylindrical Paraboloid \\
Dimensions & 305 m diameter & 104 m wide by 21 m height \\
Equivalent Dish Diameter & 305 m & 53 m \\
Frequency Range & 300 MHz to 10 GHz & 1411 to 1419 MHz (also 612 and 2650 MHz) \\
Wavelength Rage & 1 m to 3 cm & 21 cm (also 49 cm and 11 cm) \\
Illuminated Area & 213 m x 237 m (40,000 m$^2$) & $\sim$2,200 m$^2$ \\
Effective Dish Diameter & 225 m & $\sim$53 m \\
Radius of Curvature of Primary Reflector & 265 m & $\sim$243 m \\
Maximum Declination Coverage & -1 to +38 degrees & -36 to +63 degrees\\
Azimuth Pointing Range & 360 degrees & Fixed \\
Zenith Angle Range & 40 degrees & 100 degrees \\
Pointing Accuracy & $\sim$1 arcsecond & unknown \\
\enddata
\tablerefs{\citep{shuch_wow_2011,salter2012}}
\end{deluxetable*}

% Table: Compared L-band Receivers

\begin{deluxetable*}{lcc}
\tablecaption{Comparison of the observation parameters of Arecibo Wow!, relevant to this study, with those of the Ohio SETI. \label{tab:receivers}}
\tablehead{
\colhead{Property} & \colhead{Arecibo Wow!} & \colhead{Ohio SETI}
}
\startdata
Operation Years & May 2024 - Present & 1973 to 1995 (22 years) \\
Telescope & Arecibo Telescope & OSU Telescope (Big Ear) \\
Band & L-band & L-band \\
Receiver Temperature & $\sim$25 K & $\sim$100 K \\
Frequency Range & 1164.5 to 1735.5 MHz & 1420.4 to 1420.9 MHz \\
Frequency Channels & 7 x 8192 & 50\tablenotemark{a} \\
Bandwidth/channel & 10.1 kHz & 10 kHz \\
Total Bandwidth & 571 MHz & 500 kHz \\
Integration Time & 0.1 seconds & 12 seconds\tablenotemark{b} \\
Polarization & Full Stokes & Single \\
Observation Modes & ON, ON-OFF, DRIFT & DRIFT \\
Backend & Mock Spectrometer & Ohio SETI \\
Gain & 7.9 K/Jy\tablenotemark{c} & $\sim$0.1 K/Jy \\
System Equivalent Flux Density (SEFD) & 3.2 Jy\tablenotemark{c} & $\sim$1000 Jy \\
Half-Power Beam Width (HPBW) (Az x ZA) & 3.1 x 3.5 arcminutes  & 8 x 40 arcminutes \\
Transit Time (null-to-null) @ ($\pm$27° declination) & 28 seconds & 72 seconds \\
\enddata
\tablenotetext{a}{Ohio SETI operated with 8 channels, 20 to 100 kHz bandwidth, from 1973 to 1976.}
\tablenotetext{b}{The integration time was 10 seconds, plus a 2-second wait time between integrations.}
\tablenotetext{c}{Pre-Hurricane Maria values were $\sim$10.5 K/Jy and $\sim$2.4 Jy, respectively.}
\tablecomments{Az = azimuth, ZA = zenith angle.}
\tablerefs{\citep{Kraus1979, 1985IAUS..112..305D, salter2012}}
\end{deluxetable*}

\section{Arecibo SETI \label{sec:seti-arecibo}}

Arecibo's first notable involvement in SETI began in 1974 with Project Ozma II, which was a continuation of Frank Drake's original Project Ozma conducted at the Green Bank Observatory in 1960 \citep{1961PhT....14d..40D}. Project Ozma II, led by Drake himself, utilized Arecibo's then-newly upgraded 305-meter dish to search for signals from stars within 160 light-years of Earth. This project was one of the first systematic searches for extraterrestrial signals using a large radio telescope and helped establish Arecibo as a key instrument in the emerging field of SETI \citep{1975SpFl...17..421S,2021JAHH...24..981G}.

Perhaps the most famous event in Arecibo's SETI history occurred on November 16, 1974, when it was used to transmit the Arecibo Message, a 1679-bit binary message sent towards the globular star cluster M13, located approximately 25,000 light years away \citep{1975Icar...26..462.}. Designed by Frank Drake, with input from Carl Sagan and others, the Arecibo Message was a demonstration of humanity's ability to communicate with distant civilizations. Although the probability of the message being received and understood by extraterrestrial beings is slim, the event was symbolically important and underscored Arecibo's central role in the SETI community \citep{1973Natur.245..257D,1975SciAm.232e..80S}.

In the late 1980s and early 1990s, Arecibo was the primary instrument for NASA's High Resolution Microwave Survey (HRMS), a major SETI initiative \citep{1993SpFl...35..116S,1993SSRv...64...93D}. The HRMS aimed to search for narrowband signals from thousands of nearby stars, utilizing Arecibo's sensitivity and large collecting area to scan the sky more thoroughly than ever before. Although the HRMS was canceled in 1993 due to budget cuts, Arecibo's contributions during the program laid the groundwork for future SETI efforts.

Following the end of HRMS, Arecibo continued to participate in SETI through various initiatives, including the Search for Extraterrestrial Radio Emissions from Nearby Developed Intelligent Populations (SERENDIP) project \citep{1988AcAau..17..123W}. SERENDIP was a piggyback search, which means that it used data collected by Arecibo during other observations to perform a parallel search for extraterrestrial signals.

SERENDIP IV, which began in the late 1990s, used a multichannel receiver installed at Arecibo to scan the sky for signals over a wide range of frequencies \citep{2017ApJS..228...21C}. The data collected by SERENDIP IV were also used by the SETI@home project, a distributed computing effort that allowed millions of volunteers around the world to analyze SETI data on their home computers \citep{2001CSE.....3a..78K}. This collaboration between Arecibo and SETI@home significantly expanded the scope and reach of SETI research.

Another significant SETI-related effort at Arecibo was the Phoenix Project, which ran from 1995 to 2004 \citep{2003ApJS..145..181T}. Sponsored by the SETI Institute, the Phoenix Project was a targeted search that focused on specific stars thought to be good candidates for hosting intelligent life. Arecibo's large aperture and advanced receivers made it an ideal instrument for this focused search, and the Phoenix Project represented one of the most sensitive searches for extraterrestrial signals ever conducted.

Throughout its operational life, Arecibo was involved in numerous other SETI-related projects and observations, often serving as a testbed for new technologies and techniques in the search for extraterrestrial intelligence. The observatory's unparalleled sensitivity and versatility made it a cornerstone of SETI research, and its contributions continue to influence the field even after its collapse in 2020.

The Arecibo Observatory's legacy in SETI is marked by its role in pioneering early searches, conducting high-profile experiments like the Arecibo Message, and supporting a wide range of projects that advanced our understanding of the cosmos and the potential for life beyond Earth. The observatory data and findings remain valuable resources for ongoing and future SETI efforts \citep{Tarter2001}. The Big Ear and the Arecibo Observatory were two successful contemporary telescopes that also dedicated their time to SETI research (\autoref{tab:telescopes}).

% Table: Targets

\begin{deluxetable*}{lcccc}
\tablecaption{Observed targets in drift mode from the Arecibo Observatory 305-meter telescope from February to May 2020. \label{tab:targets}}
\tablehead{
\colhead{Name} & \colhead{Type} & \colhead{Distance} & \colhead{ICRS (J2000)} & \colhead{Galactic (J2000)}
}
\startdata
Teegarden's Star & dM6 Star & 3.83 pc & 02 53 00.9 +16 52 52.6 & 160.263 -37.026 \\
Barnard's Star & M4V Star & 1.83 pc & 17 57 48.5 +04 41 36.1 & 31.009 +14.062 \\
HD 157881 & K7V Star & 7.71 pc & 17 25 45.2 +02 06 41.1 & 24.758 +19.976 \\
SGR 1935+2154 & Magnetar & 9.0 kpc & 19 34 55.7 +21 53 48.2 & 57.247 +0.819 \\
3C 76.1 (B0300+162) & Radio Galaxy & 194 Mpc & 03 03 15.0 +16 26 18.9 & 163.065 -35.961 \\
3C 365 (B1756+134) & Radio Galaxy & 1500 Mpc & 17 58 24.0 +10 45 00.0 & 36.716 +16.595 \\
Void & Blank Sky & N/A & 03 00 00.0 +17 00 00 & 161.879 -35.970 \\
\enddata
\end{deluxetable*}

\section{Arecibo Wow! \label{sec:awow}} 

In 2017 we started the Arecibo REDS project (Radio Emissions from Red Dwarf Stars). This project aims to study radio emissions from red dwarf stars using data from the Arecibo Observatory. Red dwarf stars, the most abundant type in the Milky Way, emit a wide range of radio frequencies because of their high activity levels \citep{2024A&A...684A...3Y}. We are studying these emissions, focusing on their frequency, polarization, and variability. Our observations are intended to help characterize stellar flares and coronal mass ejections, which impact the space weather and habitability of their planets \citep{2020IJAsB..19..136A}.

Our research also explores the potential utility of these observations as tools for detecting exoplanets orbiting red dwarf stars \citep{2020NatAs...4..577V}. The results of Arecibo REDS will be presented elsewhere. After the collapse of the Arecibo Telescope in 2020, we plan to continue these observations with other radio telescopes.

We observed from the Arecibo Observatory's 305-meter telescope from 2017 to 2020 (AO Project A3123) and the 12-meter telescope in 2023 (AO Project A4006). Most of our targets were red dwarf stars with planets. Our observation protocol was similar to the one used by \cite{2012ApJ...747L..22R} to search for radio flares from ultracool dwarfs. Spectral observations were made in the L-band in full Stokes mode with up to 10-minute integration using the Arecibo Observatory's Mock Spectrometer. Each of the seven receiver boxes had a bandwidth of 83 MHz at central frequencies of 1282, 1353, 1424, 1495, 1566, 1637, and 1708 MHz, for a total continuous bandwidth of 509 MHz.

Similar observations were made in the S, C, and X bands. The integration was 100 ms and the channel width 10 kHz, corresponding to a Doppler change of 2.1 km/s relative to the \HI line (unintentionally the same as the Big Ear during the \wow detection). There were 8192 channels per receiver. We used both diode and target calibrators to calibrate our flux density measurements.

Arecibo Wow! is a technosignatures side project of of Arecibo REDS. Our original observational strategy was neither tailored nor optimized for the detection of technosignatures, e.g. a high time and frequency resolution \citep{wright2021strategies}. However, our sample included stars with known potentially habitable exoplanets \citep{hwc2024}, which were observed for long sessions that spanned frequencies from 1 to 10 GHz.

In this study, we focus only on our drift-scan mode observations made from February to May 2020 at topocentric frequencies between 1419.5 and 1421.0 MHz (\autoref{tab:suppl-scans}). This mode and frequency range are analogous to the protocol used in the detection of the \wow (\autoref{tab:receivers}).

The seven targets studied with drift scans were Teegarden's Star, Barnard's Star, and HD 157881, which were part of our main study (\autoref{tab:targets}). We also conducted observations in this mode of the radio galaxies 3C 76.1 (B0300+162) and 3C 365 (B1756+134), which served as our calibration standards. The magnetar SGR 1935+2154 constituted an opportunistic target, due to its active state during the observation period and its proximal spatial alignment with our primary subjects \citep{2020Natur.587...59B}. Our final target was a nearby random section of the sky that we denoted Void, for reference purposes.

\section{Data Analysis and Results \label{sec:analysis}} 

\begin{figure}
    \centering
    \includegraphics[width=1.0\linewidth]{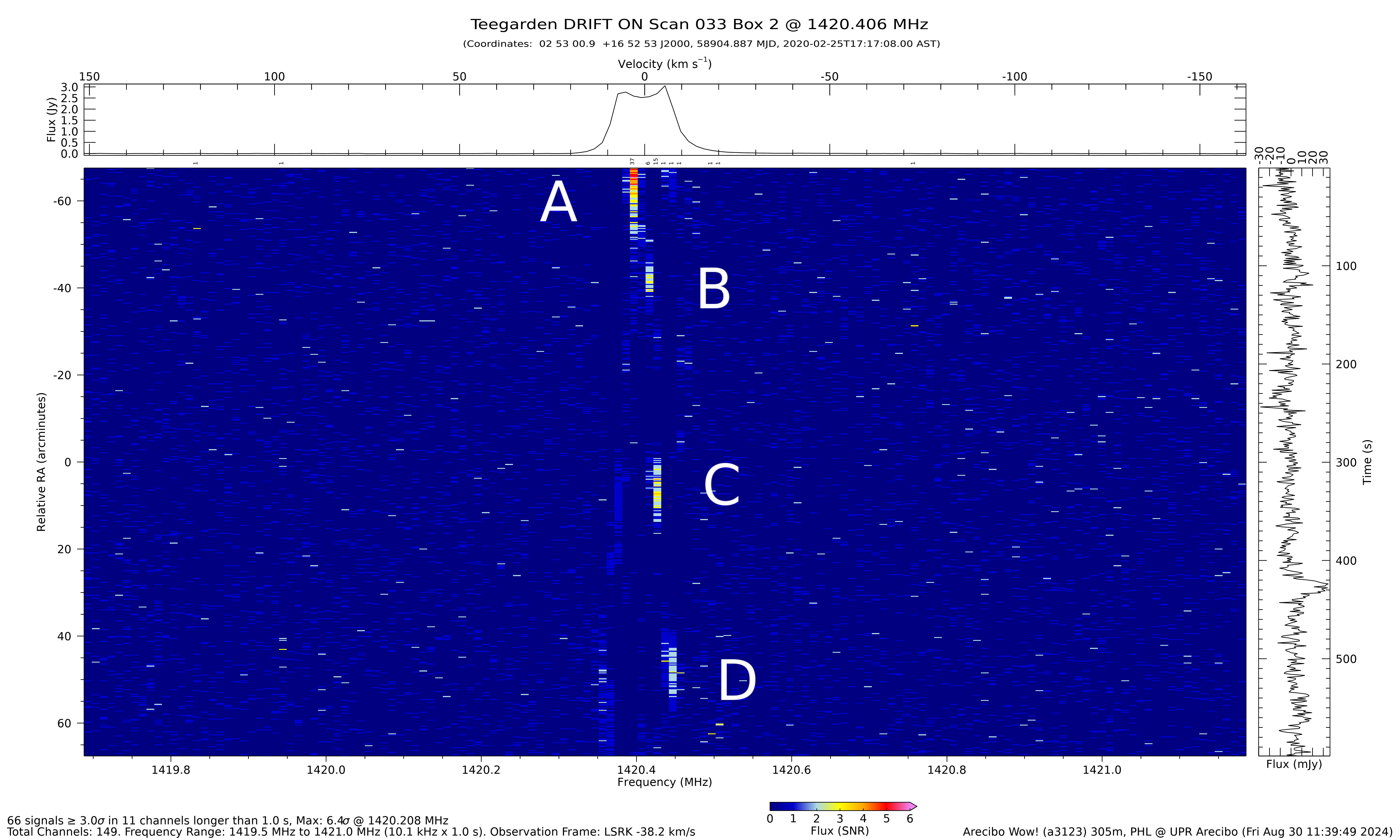}
    \caption{Dynamic spectra of a 10-minute drift scan centered on Teegarden’s Star showing four narrowband signals associated to small cold \HI clouds, labeled A, B, C, and D. The top subplot is the time average flux density with the galactic \HI profile at the center. The right subplot is the frequency average flux density with a peak between 400 and 500 seconds attributable to AGC 74887. The clouds are not related to Teegarden’s Star or AGC 74887. Their properties are described in \autoref{tab:clouds}. \label{fig:teegarden}}
\end{figure}

The data were read and calibrated using standard procedures with the Arecibo Software Library. The dynamic spectra were detrended and normalized in frequency and time to cancel out broadband signals and identify narrowband variations. Most of the observations were made in the LSRK frame, but were also converted to the topocentric frame to make it easier to identify terrestrial RFI. Our longest drift scan was 10 minutes long (2.5 degrees in right ascension), of Teegarden's star (\autoref{fig:teegarden}).

% Figure: Dynamic Spectra of Teegarden

We developed a series of methodologies in the Interactive Data Language (IDL) aimed at the detection of narrowband signals. Among these, one emulated the analytical approach employed by the Ohio SETI, specifically averaging data over 10-second intervals and using the most recent 60 data points for mean and background noise calculation. The flux density was then converted to a signal-to-noise ratio (SNR). All of the methods yielded consistent results.

We identified four main signals in the vicinity of Teegarden's Star corresponding to small cold neutral hydrogen (\HI) clouds (\autoref{fig:teegarden}). Similar clouds appeared in the observations of SGR 1935+2154 and in the Void region. The intensity and location of the cloud signals were consistent between multiple independent scans during the same observation session of Teegarden's Star. There was some variability in the SGR 1935+2154 signals observed in three epochs separated by a week, which could be attributed to noise or scintillation.

As an example, we focus our analysis on the clouds near Teegarden's Star, since it was the longest drift scan we performed. Their observed properties are consistent with small cold \HI clouds (\autoref{tab:clouds} and \autoref{fig:suppl-clouds}). Their hydrogen column density $n_{HI}$ corresponds to their partial contribution to the total \HI background (\autoref{fig:nhi}). See the supplementary \autoref{fig:suppl-4HPI} and \autoref{fig:suppl-GALFA} for a description of $n_{HI}$ in all our observed targets.

\begin{figure}
    \centering
    \subfigure[Teegarden's Star (GALFA).]{
        \includegraphics[width=1\linewidth]{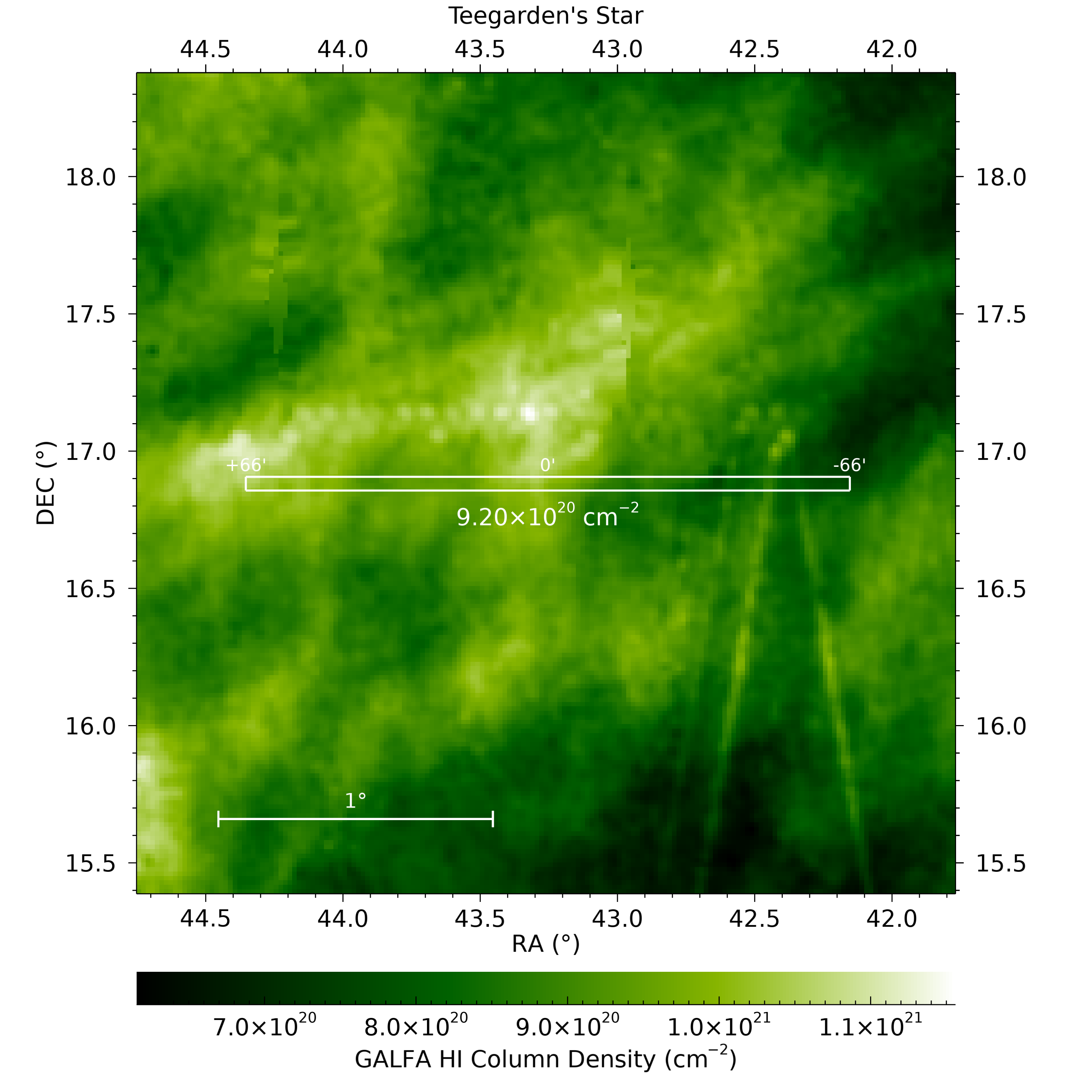}
    }
    \subfigure[\wow (GALFA).]{
        \includegraphics[width=1\linewidth]{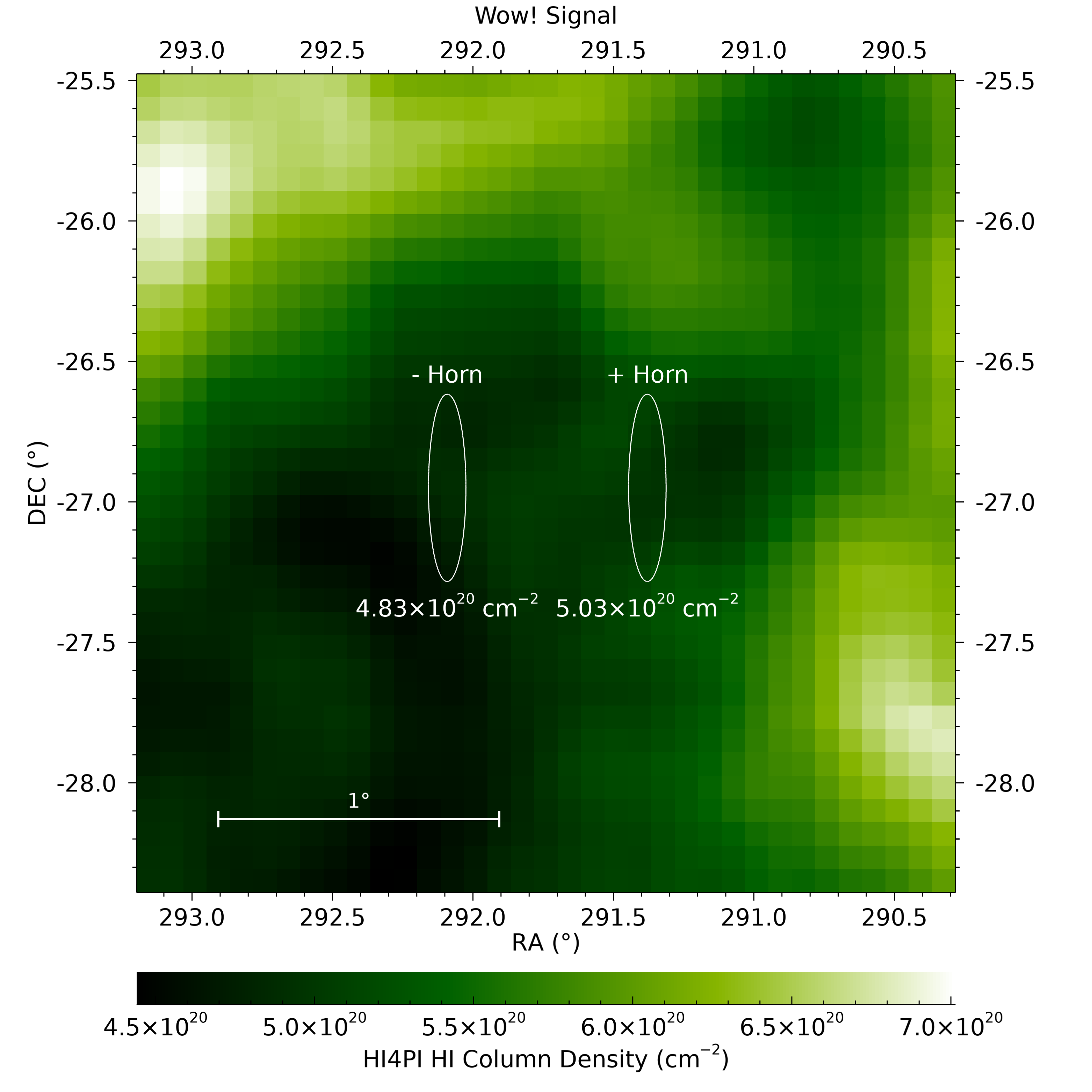}
    }
    \caption{Hydrogen column densities $n_{HI}$ near the Teegarden's Star and the \wow fields, figures (a) and (b), respectively. The scan strip of our observation and its average density are shown (white frame and text). The \wow field shows its two possible location with a lower $n_{HI}$. Note that the colorbar does not have the same scale in each plot to highlight the relative column density. Data from the Galactic Arecibo L-Band Feed Array (GALFA) and the \HI 4$\pi$ (HI4PI) surveys, with spatial resolution of 4 and 16 arcminutes, respectively \cite{2016A&A...594A.116H,2018ApJS..234....2P}.\label{fig:nhi}}
\end{figure}

The most noticeable characteristic of the \HI cloud signals depicted in \autoref{fig:teegarden}, is their close resemblance to the bandwidth and duration of the \wow (\autoref{fig:printout}). They are concentrated primarily in a bandwidth of $\approx$ 10 kHz, but it is clear that this is an artifact of the frequency resolution, and their actual bandwidth could be smaller or even slightly larger. These clouds are spatially resolved, among the characteristic galactic \HI profile, with different frequencies due to different Doppler velocities.

The angular dimensions of these \HI clouds are comparable to the azimuth beam width of the Big Ear telescope, approximately 8 arcminutes (\autoref{tab:receivers}). Their brightness might fall within or below the detection sensitivity of the Big Ear. The primary distinction between these signals and the \wow is their brightness and transient characteristics, that is, they have to become about one-hundred times brighter for a few minutes to reproduce the \wow.

% Table: Teegarden Clouds

\begin{deluxetable*}{ccccccccc}
\tablecaption{Observed properties of the small neutral hydrogen clouds, labeled A, B, C, and D, in the field vicinity of Teegarden's Star. These clouds have a kinetic temperature lower than 97 K. Their linear and circular polarization is less than 10$\%$ and 3$\%$ (3$\sigma$), respectively. Cloud A is not well constrained and the presented values are upper limits.\label{tab:clouds}}
\tablehead{
\colhead{Cloud} & \colhead{RA} & \colhead{$\nu$ (MHz)} & \colhead{v (km/s)}& \colhead{$S_\nu$ (mJy)} & \colhead{LPol ($\%$)} & \colhead{T$_b$ (K)} & \colhead{$\theta$ (arcmin)} & \colhead{N$_{HI}$ (cm$^{-2}$)}
}
\startdata
A & 02h 59m 18s & 1420.382 & +5.1 & $\le$2077 $\pm$ 242 & $\leq$21.4 & $\leq$52 & $\leq$60.3 & $\leq$2.02$\times10^{20}$ \\
B & 02h 55m 19s & 1420.402 & +0.9 & 387 $\pm$ 24      & 5.4       & 9      & 11.6    & 3.77$\times10^{19}$ \\
C & 02h 52m 04s & 1420.412 & -1.3 & 501 $\pm$ 28      & 7.0        & 12      & 12.3    & 4.88$\times10^{19}$ \\
D & 02h 49m 16s & 1420.432 & -5.5 & 526 $\pm$ 26      & 9.9        & 13      & 15.7    & 5.13$\times10^{19}$ \\
\enddata
\tablecomments{All clouds share the same +16d 53m 00.9s declination.}
\end{deluxetable*}

\section{Proposed Mechanisms \label{sec:mechanism}}

We postulate that when observed by the Big Ear telescope in 1977 one of the small cold \HI clouds in the beam of the telescope experienced a transient increase in brightness because of a natural astrophysical process. The cloud itself was not bright enough for Big Ear to detect on its own outside of this event. In other words, the \wow was a natural phenomenon that was observed unexpectedly when a radio telescope was aimed at the right place and time in the sky.

Several known mechanisms could account for a transient surge in the brightness of the hydrogen line (see \textit{e.g.}, \citealt{wouthuysen1952excitation,mckee1977theory,lazarian2000velocity,liszt2001spin, furlanetto2006cosmology}); however, most of such processes are also likely to increase the thermal energy or kinematic activity of hydrogen within the interstellar medium, thus broadening the \HI line beyond the observed 10 kHz range. Moreover, an increase in spontaneous emission will hardly drive the large surge in brightness able to explain the observed flux in the \wow (see next section).

Before assessing the astrophysical mechanisms that can explain the peculiar characteristics of the \wow, we need first to constrain some of the properties of the hypothetical source. For this purpose, we will assume, as suggested by our data, that the source of the signal was a small \HI cloud in the cold neutral medium (CNM).

\subsection{Astrophysical Properties of the \wow \label{subsec:astrowow}}

If we assume that the \wow was produced by some unknown astrophysical mechanism involving \HI clouds, we need to fulfill several conditions in order to explain some of their characteristics. A signal coming from a cloud with kinetic temperature $T_K$ will have Doppler broadening \citep{rybicki1991radiative}:
\begin{equation}
\Delta \nu_{\mathrm{FWHM}}=\sqrt{\frac{8 k_B T_K \ln 2}{m c^2}} \nu_{10}
\end{equation}
where $k_B$ is the Boltzmann constant, $m$ is the average mass of the gas particles, and $\nu_{10}=1420.406$ MHz is the frequency of the photon emitted after the transition between the two hyperfine states of the ground hydrogen atom, \textit{i.e.}, the 21-cm hydrogen line. In order to have an emission with $\Delta \nu_\mathrm{Wow!}\leq10$ kHz, the kinetic temperature should not be greater than:
\begin{equation}
T_K \leq T_{\mathrm{Wow!}} = \frac{mc^2}{8\ln 2 k_B}\left(\frac{\Delta \nu_\mathrm{Wow!}}{\nu_{10}}\right)^2 = 97 K.
\end{equation}
This temperature threshold is not larger than the kinetic temperature of diffuse clouds in the interstellar medium (ISM) \citep{cox2005three}, which are precisely the proposed sources of the signal. 

According to Ohio SETI data, the signal had an almost constant intensity for at least 72 seconds \citep{ehman2010big}. Since the detector stayed at the same declination for almost 60 days and no signal with comparable intensity was detected 24 hours later, we know that the duration of the signal was not greater than $24\;\mathrm{h} \approx 8.6\times 10^4$ s, at least after signal detection. This is consistent with the recent search for repetitions that ruled out a maximum duration of 40 hours \citep{gray2002search,kipping2022could}. 

On the other hand, the \wow was detected at a frequency of $\nu = 1420.4556$ MHz \citep{harp2020ata}, very close to the central hyperfine transition frequency $\nu_{10}$. This corresponds to a relative velocity of only 10.5 km/s. This fact has been used repeatedly as an argument in favor of an astrophysical origin of the signal. However, given the low galactic longitude, namely $l=11.7^\circ$, a low radial velocity will not be rare among nearby sources. A first-order estimation to the distance of a galactic source having a given radial velocity with respect to the Sun, can be obtained from the Oort's constants and other galactic parameters using \citep{wang2021local}:
\begin{equation}
V_r = d(K+A \sin 2 l+C \cos 2 l) \cos^2 b 
\label{eq:vr}
\end{equation}
where $A=16.31 \pm$ $0.89 \mathrm{~km} \mathrm{~s}^{-1} \mathrm{kpc}^{-1}, B=-11.99 \pm 0.79 \mathrm{~km} \mathrm{~s}^{-1} \mathrm{kpc}^{-1}, C=-3.10 \pm 0.48 \mathrm{~km} \mathrm{~s}^{-1} \mathrm{kpc}^{-1}$, and $K=-1.25 \pm 1.04 \mathrm{~km} \mathrm{~s}^{-1} \mathrm{kpc}^{-1}$ are the Oort constants corresponding to a non-axisymmetric disk. Since we know that $v_\mathrm{r,Wow!}\leq 10$ km/s, we estimate, from pure kinematical arguments, that the maximum distance to the source should be $d_\mathrm{Wow!}\leq 4.9\pm 1.8$ kpc, where the error in the distance comes from uncertainties in the Oort constants. In fact, a minimum value of $2$ kpc is compatible with a radial velocity of 10 km/s.  

A further and tighter constraint on the distance, can be obtained from the vertical scale height of the \HI distribution. Using data from the Leiden-Argentine-Bonn survey \citet{kalberla2009hi} estimated typical scale heights at the solar distance, between 150 pc and 400 pc for the cold and warm components of the diffuse neutral ISM. This implies that at the galactic latitude of the \wow, $b = -18.9^\circ$, the maximum distance of an ISM filament will be between 0.5 and 1.2 kpc. These values are compatible with the kinematical constraint. However, the lower value of the distance, namely 0.5 kpc, would imply relative radial velocities of a maximum of 3 km/s.

If the \wow was produced in the diffuse cold ISM, it must be a relatively local phenomenon. Either it is relatively high above the mean galactic disk, or it is close by but has a peculiar radial velocity of around 7 km/s as compared to its stellar environment. 

Using the previous constraint on the distance, we can now estimate the isotropic luminosity of the signal. The total intensity recorded by Ohio SETI was in the range of $54-212$ Jy (\autoref{tab:wow}). Using three characteristic distances of 0.5, 1.2 and 4.8 kpc, and assuming the lower more conservative value of 54 Jy, the total isotropic luminosity of the signal could be $L_\mathrm{iso} = 4.2\times 10^{-8}, 2.4\times 10^{-7}$ and $3.9\times 10^{-6}\;L_\odot$, respectively.

In summary, according to our observational evidence, as well as previous constraints, the astrophysical process able to explain the \wow should have the following key properties: 1) it must be narrowband, 2) transient, with a duration of the order of minutes to hours; and 3) relatively luminous, with maximum isotropic luminosities of the order of $10^{-8}-10^{-6}\;L_\odot$. 

\subsection{Superradiance}

The most suitable astrophysical mechanism capable of producing intense narrowband emission, at short time scales, is the so-called superradiance, hereafter SR. This is the generic name of a process in which the emission of radiation from an optically thin source depends non-linearly on the density of the emitting medium \citep{rajabi2016dicke}.

There are two types of astrophysical SR. There is spontaneous superradiance (SSR), which is produced when molecules or atoms in a medium are pumped into excited meta-stable states, a process called {\em population inversion}; the pumping is achieved by incoming radiation, shocks, and/or collisions. With the proper stimuli and environmental conditions, \textit{e.g.}, coherent gas motions, large column densities, among others, stimulated emission of radiation can exponentially induce the emission of radiation from the medium, attaining some degree of spatial coherence (beaming), narrowband emission, and relatively large luminosities. Spontaneous SR is complementary to {\em astrophysical masers}, leading to intense radiation bursts in maser-hosting regions. 

On the other hand, we have coherent SR, also known as Dicke's superradiance (DSR) after the seminal work of \citet{dicke1954coherence}. In DSR, tightly packed atoms in a populated inversion state would radiate coherently, as opposed to spontaneously, due to the connection among their states induced by their common electromagnetic field. In the following sections, we discuss the requirements for an SSR and DSR phenomenon to occur.

\subsection{SSR: The \wow as a Maser Flare \label{subsec:maser}}

Astronomical hydrogen masers are rare. The first one was discovered in 1989 around the star MWC349 \citep{1989A&A...215L..13M}. Its emission exhibited characteristics similar to those of a molecular maser source. It was highly luminous and fluctuated over time, because of its sensitivity to alterations in the intricate excitation processes. However, hydrogen masers operating at the \HI line frequency have yet to be detected in astronomical observations, though they have been produced in laboratory settings \citep{major2007hydrogen}. In fact, human-made hydrogen masers are used for timekeeping in astronomical observatories \citep{gray_maser_2012}. 

The first serious proponents of the existence of \HI masers were the Russian astrophysicists I.S. Shklovskii and D.A. Varshalovich in 1967. They intended to introduce the maser action in the 21-cm line to explain the discrepancy between the observed ISM neutral hydrogen density and the rate of accretion in the galactic disk \citep{storer1968interstellar}. Although the aforementioned discrepancy was solved by other astrophysical processes, the basic \HI pumping mechanism devised by the Russian pioneers remains the most suitable way to produce population inversion in a neutral hydrogen region.

The pumping mechanism devised by Shklovskii and Varshalovich involves the absorption of Lyman $\alpha$ photons from nearby \HII regions (see Figure 1 in \citealt{storer1968interstellar} and the detailed explanation there in). Consequently, if an \HI region is subject to a sudden increase in Ly$\alpha$ photons, the conditions for the maser action could arise. 

The maser action exponentially amplifies the intensity of the incoming radiation $I_\nu(0)$, when it traverses a plane-parallel medium having a spatial thickness $L$ according to \citep{gray2012maser}:
\begin{equation}
I_\nu(z) = I_\nu(0) \exp(\gamma_\nu L). 
\end{equation}
Here, $\gamma_\nu$ is the so-called gain factor. At the center of the 21 cm line, the gain factor is \citep{storer1968interstellar}:
\begin{equation}
\gamma_0=\frac{3}{32 \pi \nu_{10}^2 \Delta \nu} c^2 A_{10} n_{\mathrm{H}} f = 3.8\times 10^{-14}\left(\frac{n_H}{100\;\mathrm{cm}^{-3}}\right) f
\end{equation}
where $A_{10}=2.85\times 10^{-15}$ s$^{-1}$ is the Einstein A coefficient, which gives us the probability of the $1\rightarrow 0$ transition,  $n_{\mathrm{H}}$ is the neutral hydrogen density and $\Delta\nu\approx 10$ kHz is the Doppler broadening. The factor $f$ is called the degree of inversion, and it is given by:
\begin{equation}
f=\frac{g_0 n_1}{g_1 n_0} - 1\approx \exp\left(-\frac{h\nu_{10}}{k_B T_s}\right) - 1\approx -\frac{6.81 \times 10^{-2}\;\mathrm{K}}{T_s}
\end{equation}
where $T_s$ is called the spin temperature \citep{field1958excitation}.

If the medium is irradiated with a pumping (Ly$\alpha$) photon field, number density $n_\nu\;d\nu$, the spin temperature will be given by \citep{storer1968interstellar}:
\begin{equation}
T_s\approx -\frac{6.81 \times 10^{-2}\;\mathrm{K}}{\nu_{10}} \frac{n(\nu)}{\;dn(\nu)/d\nu}
\end{equation}
and hence the degree of inversion will be:
\begin{equation}
f = \nu_{10}\frac{dn(\nu)/d\nu}{n(\nu)}
\end{equation}
\citet{storer1968interstellar} showed that the maser action when the radiation field is such that: 
\begin{equation}
\frac{dn_\nu}{d\nu}>2\times 10^{-26}\;\mathrm{photons\; Hz}^{-2}\;\mathrm{m}^{-3}.
\end{equation}
With these equations at hand, we can evaluate, under a given circumstance, whether a region of the ISM will produce a 21-cm maser, \textit{i.e.}, if $\gamma_0 L\gg 1$.

In the original mechanism proposed by Shklovskii, Varshalovich, Storer, and Sciama, the source of Ly$\alpha$ photons is a nearby \HII region. In fact, the conditions for population inversion and therefore 21-cm maser emission were identified in the Orion's Veil \citep{abel2006physical}. However, in the case of the \wow, none of the \HII regions, such as Sagittarius A (Sgr A) or Sagittarius B2 (Sgr B2), are known in the sky near the signal location. Other sources of Ly$\alpha$ photons should be found. 

In particular, if an intense event occurs in the vicinity of the cloud, producing continuous radiation, we can estimate the amount of the right photons arriving at the cloud. Let us assume that the high-energy event produces at the frequency of interest a black body spectra with a brightness temperature $T_b\gg h\nu_{Ly\alpha}$, from a characteristic radius $R$ and it is located at a distance $d$. The photon number density will be:
\begin{equation}
n_\nu(\nu, T)=\frac{8 \pi \nu^2}{c^3} \frac{1}{\exp \left(\frac{h \nu}{k_{\mathrm{B}} T_b}\right)-1}\frac{R^2}{d^2}\approx \frac{8 \pi \nu}{h c^3} k_B T_b\frac{R^2}{d^2}  
\end{equation}
and hence:
\begin{equation}
\frac{dn_\nu}{d\nu} = \frac{8 \pi}{h c^3} k_B T_b\frac{R^2}{d^2}.
\end{equation}
Interestingly, in this case the degree of inversion will depend only on the ratio of the 21-cm to the Ly$\alpha$ frequency:
\begin{equation}
f \approx \frac{\nu_{10}}{\nu_{\mathrm{Ly}\alpha}} \approx 5.8 \times 10^{-7} 
\end{equation}
which correspond to an almost total inversion of the population. 

Now we need to ask if the radiation field is strong enough for the maser action. To ensure this, we need to guarantee that:
\begin{equation}
\frac{dn_\nu}{d\nu}=\frac{8 \pi}{h c^3} k_B T_b\frac{R^2}{d^2}>2\times 10^{-26},
\end{equation}
which can be more conveniently expressed as:
\begin{equation}
\left(\frac{T_b}{10^{10}\;\mathrm{K}}\right) 
\left(\frac{R}{30\;\mathrm{km}}\right)^2 
\left(\frac{0.1\;\mathrm{pc}}{d}\right)^2
> 1.
\end{equation}
We can see that only if the source has a very large brightness temperature and is relatively close the inversion condition is fulfilled. On the other hand, the gain factor is: 
\begin{equation}
\gamma_0 L = 2.7\left(\frac{n_H}{4\times 10^5\;\mathrm{cm}^{-3}}\right)\left(\frac{L}{1\;\mathrm{pc}}\right),
\end{equation}
which implies that in order to have a maser effect the density of hydrogen should be 1000 times larger than in the diffuse clouds and its extension should be of several parsecs.

\subsection{DSR: The \wow as a Dicke's Superradiance Event \label{subsec:dsr}}

DSR arises when a coherent emission of radiation is induced in a population-inverted medium. The mechanism is physically analogous to stimulated superradiance and similar to a maser, but the quantum mechanical nature of the phenomenon, which involves the entanglement of many atoms connected by the electromagnetic fields of their own radiation, give it some distinctive properties particularly well suited for the purpose of this work.

DSR was first predicted and described by \citet{dicke1954coherence}. Since this seminal work, the phenomenon has been extensively studied in the literature and more importantly in the laboratory (for a complete review, see \citealt{gross1982superradiance}). In recent years, it has attracted the attention of the astrophysical community \citep{rajabi2016dicke,houde2018explaining,rajabi2020astronomical,houde2022variability}. In what has also become a seminal paper in the area \citet{rajabi2016dicke}, hereafter RH2016, applied the DSR formalism to study the conditions and properties of coherent superradiance in neutral hydrogen sources. Here, we adapt some of their results to the context of the \wow.

\begin{figure}
    \centering
    \includegraphics[width=1\linewidth]{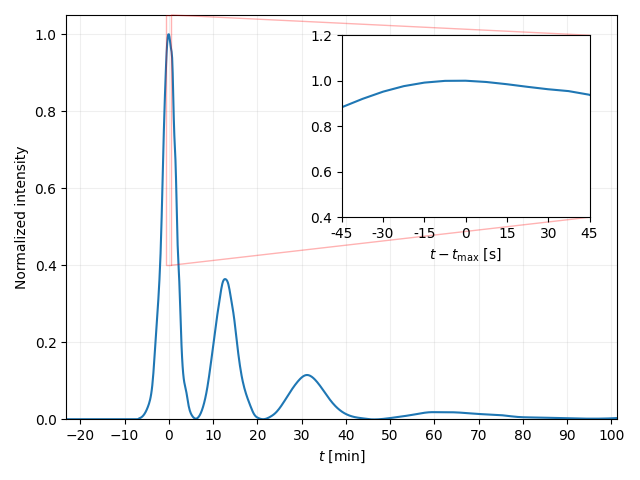}
    \caption{Normalized intensity as a function of time for the region in the example of the text. \label{fig:DSR}}
\end{figure}

The most interesting aspect of DSR is the beaming and time structure of the signal. In \autoref{fig:DSR} we plot the normalized radiation intensity for one of the systems studied in RH2016. 

%The DSR mechanism is particularly well suited for inducing intense radiation emission along elongated, pencil-like or cylindrical structures. The system whose superradiance is represented in \autoref{fig:DSR} correspond to a cylinder fill with \HI with a number density $n_H = 10$ cm$^{-3}$, inversion factor $\eta = 0.01$ and having a total length $L=2\times 10^{-6}$ pc and a diameter $w=3\times 10^5$ m.

The DSR mechanism is particularly well suited for inducing intense radiation emission along elongated, pencil-like, or cylindrical structures. The system whose superradiance is represented in \autoref{fig:DSR} corresponds to a cylinder volume of \HI with a number density $n_H = 10$ cm$^{-3}$, inversion factor $\eta = 0.01$ and having a total length $L=2\times 10^{11}$ m and a diameter $w=\sqrt{\lambda_{10}L/\pi} = 1.2\times 10^5$ m.\footnote{We scaled down the original result published in \citet{rajabi2016dicke} to a smaller emission region. As a result, the characteristic time of the superradiance is increased by a factor of 5. This does not affect the conditions required to have superradiance.}

%In the original paper, the region has $L=10^{12}$ and $w=9\times 10^{5}$ cm. This does not affect the conditions required to have superradiance.}

DSR emission is characterized by three time scales. In all models the population-inverted system must build some degree of correlation before producing superradiance. The time required for this accumulation process is called the {\em delay time} $t_D$. Once superradiance is achieved, the total emission time is quantified by the {\em characteristic time of superradiance} $T_R=16 \pi \tau_{\mathrm{sp}}/(3 n \lambda^2 L)$, where $n$ is the number density of inverted atoms and $\tau_{\mathrm{sp}}$ is the spontaneous decay time. In the example in \autoref{fig:DSR}, $T_R\approx 5$ s. Finally, since the cloud has a finite size, it will take a time $\tau_E=L/c$ for the radiation to propagate along its largest axis. The finite-time propagation is partially responsible for the ringing structure of the signal. In the example case considered here $\tau_E = 11$ min. 

Observing the total time of the DSR emission from the hypothetical region studied here, we can understand how a signal lasting for a couple of minutes may arise from a natural astrophysical phenomenon and then fade render it undetectable even a few hours after its observation. 

RH2016 has estimated that the integrated flux of the DSR emission produced by the region in \autoref{fig:DSR} will be as low as $\sim 10^{-25}$ W m$^{-2}$ if located at a distance of 0.4 kpc, which is close to the lowest estimated distance for the \wow (see \autoref{subsec:astrowow}). Assuming that this flux is uniformly emitted in a band as narrow as that associated with its temperature, the corresponding spectral flux density will be $\sim 1$  mJy. However, if, as it was also suggested in RH2016, we have a large number of superradiant small regions that erupt around the same time, the observed spectral flux could be considerably larger. 

\begin{figure}
    \centering
    \includegraphics[width=1\linewidth]{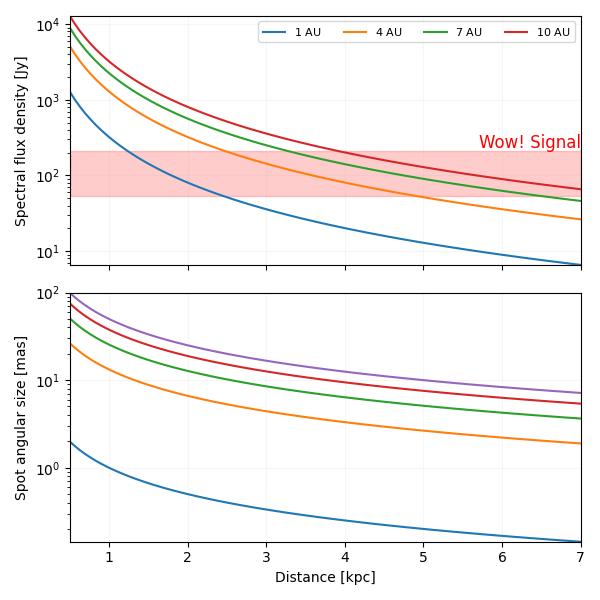}
    \caption{Spectral density and spot angular size of a superradiant region with the properties described in the text, as a function of region size and distance. The intensity of the \wow can be explained for a \HI cloud located at around 0.5 kpc. If the source was much further, a DSR emitting region as large as 50 AU will be required. In all cases the size of the region will be hard to resolve. \label{fig:IDSR-distance}}
\end{figure}

If we assume, for instance, that coherent DSR occurs over a spot with a size $D_s$, the total spectral flux at a observed from a distance $d_{Wow!}$ will be:
\begin{equation}
I_\mathrm{DSR} < 1\;\mathrm{mJy} \left(\frac{D_s}{w}\right)\left(\frac{0.4\;\mathrm{kpc}}{d_s}\right)^2.
\end{equation}
In \autoref{fig:IDSR-distance} we plot the spectral flux density as a function of spot size and distance. We verify that a DSR with the properties assumed in RH2016 is able to explain not only the duration but the observed flux of the \wow. However, in all cases, the angular size of the spot is much smaller than the resolution of past and even current detectors. 

% Figure: DSR flux density

% Figure: Geometry

\begin{figure*}
    \centering
    \includegraphics[width=1\linewidth]{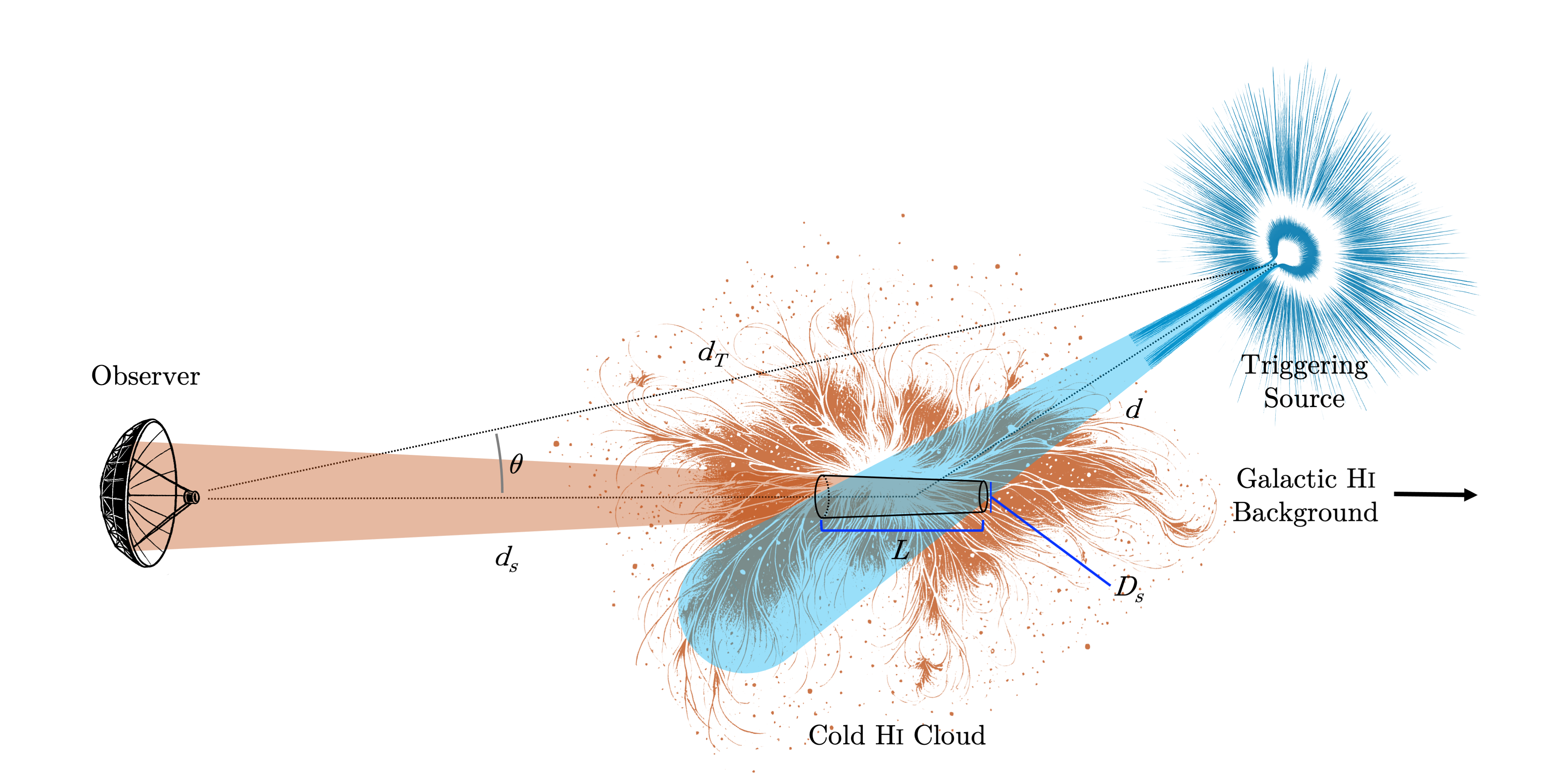}
    \caption{The proposed \wow emission source is a region of a cold \HI cloud at a distance $d_s$ that emits a superradiance radio beam along a line of sight $L$ and with a diameter $D_s$. This event is triggered by a strong radiation source, at a distance $d$ from the cloud and $d_T$ from the observer. The trigger beam is not necessarily observed depending on its distance, size, and separation angle $\theta$. Since the superradiance event also takes a time to build up, the trigger beam if observed, always precedes the superradiance event by seconds to hours. \label{fig:cloud}}
\end{figure*}

\section{Discussion \label{sec:discussion}}

We postulate that the \wow was a superradiance event that produced a maser-like flare from a small cold \HI cloud (\autoref{fig:cloud}). The trigger source was intense enough in the radio spectrum to saturate or invert a large fraction of the \HI atoms out of equilibrium. Thus, a magnetar flare could provide the required photon intensity to trigger a narrowband superradiance event in the \HI line. However, we cannot rule out other sources of radiation as long as they do not ionize or introduce additional kinetic energy to the cloud.

Magnetars are neutron stars with extremely strong magnetic fields. They have become increasingly recognized for their role in explaining various astrophysical transients. Young millisecond-period magnetars can release their spin-down energy to power bright phenomena like Gamma-ray Bursts (GRBs), Super-Luminous Supernovae (SLSNe), and fast X-ray transients such as CDF-S XT-2 \citep{2021ApJ...915L..11A}. Older magnetars are known to produce diverse transients by expending their magnetic energy, resulting in giant flares, soft gamma repeaters (SGRs) or hard X-ray bursts, and occasionally fast radio bursts (FRBs). Some magnetar giant flares have even been detected as short GRBs from nearby galaxies \citep{2022arXiv220507670Z}.

There are three plausible explanations for the lack of detection of the triggering source, \textit{e.g.}, a magnetar flare, and only the emission from the \HI cloud being observed by Ohio SETI. First, the source may be insufficiently intense to be detected by the telescope but still adequate for amplification. For example, the intensity received in the \HI cloud could be substantially higher if the source is in proximity or within the cloud.

Second, despite the almost instantaneous nature of stimulated emission, where the source's broadband signal and the amplified emission in the \HI line should technically arrive simultaneously, there may be a temporal delay attributed to processes of absorption and re-emission. In scenarios involving a large population of atoms, achieving maximum amplification requires a short but finite time frame, such as DSR, potentially resulting in a signal delay ranging from seconds to minutes.

Third, contrary to masers, the alignment of the trigger and cloud beam do not have to be parallel. Thus, the trigger beam is not necessarily visible from Earth, depending on its size and orientation. For example, theoretical models suggest magnetar beams 2° to 6° wide, but they might have a complex structure \citep{2016MNRAS.461..877V}. However, we believe that the most favorable alignment is for the source and the cloud to be close to the line of sight or far behind the cloud, \textit{i.e.}, with a small angular separation in the sky. 

Small cold \HI clouds are significant components of the ISM, often serving as precursors to molecular clouds. Recent studies reveal that these clouds can exhibit unexpectedly high column densities and complex structures, challenging previous assumptions about their abundance and characteristics. They can reach column densities of $\geq$10$^{22}$ cm$^{-2}$ in molecular clouds, with temperatures as low as $\approx$10 K. This suggests that previous estimates of cold \HI mass may be underestimated by a factor of 3 to 10 due to observational biases such as \HI self-absorption (HISA) \citep{david_j__grynkiewicz_2022}.

Observations of the Riegel-Crutcher cloud revealed a network of thin, elongated filaments of cold HI, with individual strands measuring up to 17 pc long and only 0.1 pc wide \citep{2007ASPC..365...65M}. This structure supports theories of the Tiny Scale Atomic Structure (TSAS) in the interstellar medium \citep{heiles1997tiny,2003ApJ...598L..23S}. The Arecibo Telescope has been instrumental in studying TSAS, with observations indicating that although TSAS can be detected on scales as small as 10 AU, they are relatively rare in the ISM \citep{2010ApJ...720..415S}.

It is interesting to note that the Ohio SETI identified cold hydrogen clouds in the first months of its project \citep{1977Icar...30..267D}. The signals of these clouds were analogous to the narrowband (then 20 kHz) signals pursued during the initial phase of the project (eventually refined to 10 kHz). They were identified as small-diameter cold hydrogen clouds because they were shown to be continuum sources that exceeded the beamwidth.

Subsequently, observations with the Very Large Array (VLA) uncovered many small unidentified continuum sources within the \wow field that were below the detection threshold of the Big Ear telescope \citep{gray_vla_2001}. All these sources were excluded as the origin of the \wow due to the absence of a transient narrowband mechanism capable of increasing their flux by two orders of magnitude.

The exact location of the \wow remains unknown. The Big Ear radio telescope was characterized by dual beamwidths, termed negative (first or West) and positive (second or East) horns, each encompassing many stellar objects. We believe that the most plausible position of the source is within the second horn, owing to its location with a higher and complex column density (\autoref{fig:nhi}) and more potential sources in the VLA data (Figure 2 of \cite{gray_vla_2001}). 

Furthermore, the attenuation of astrophysical signals typically extends significantly longer relative to their rise. Consequently, a signal initially observed in the first horn was more likely to appear in the second horn after a lapse of three minutes, consistent with its decay. Hence, a signal detected in the second horn represents a more plausible scenario, further supported by the \HI column density and the VLA data.

Potential candidates for the precise location of the \HI cloud region associated with the \wow could be identified following additional observations. The \HI lines of these candidates should have a bandwidth $\leq$10 kHz, a 10 km/s blue shift relative to the \HI line, and a size of less than 8 arcminutes (\textit{i.e.} the azimuth beamwidth of the Big Ear), although the emission region within the cloud could be much smaller. 

We can also infer that the flux density of this \HI cloud was below the detection limit of the Big Ear, but not necessarily observable by current instruments. We believe that it was not a coincidence that the source of the \wow exhibited a dynamic correlation with the motion of the galactic \HI gas \citep{gray_search_2002}. This correlation also supports our hypothesis that the signal was an astrophysical event associated with \HI clouds.

Investigating these \HI clouds is more effectively done using a large radio telescope with good spatial resolution, such as the VLA or China's Five-hundred-meter Aperture Spherical Telescope (FAST). Unfortunately, the identification of the triggering source is more challenging, assuming, for example, that the event was attributed to an unknown magnetar. Flaring events from magnetars are infrequent and must be directly aligned towards Earth for detection. For example, it is not possible to tell if the closest magnetars to the \wow field, SGR 1806-20, SGR 1900+14, or SGR J1745-2900 were responsible.

Persistent monitoring of cold \HI clouds and similar regions for \textit{wow signals} may reveal recurrent events. However, these observations face intrinsic difficulties due to the transient nature and rarity of the event, which may explain the unique nature of the \wow to date. Another option is to search for \textit{wow signals} in archive data. The Ohio SETI program is the longest-lasting SETI program, spanning 22 years of continuous operation. Its extended survey may explain their fortuitous detection of this event. 

% We have a "prediction with no a priori falsifiability" or a "verifiable prediction".

\section{Summary and conclusions \label{sec:conclusion}}

We observed from the Arecibo Observatory narrowband signals ($\Delta\nu\leq$ 10 kHz) similar to the \wow near the hydrogen line, though significantly weaker. These signals are attributed to small cold \HI cloud regions within the galactic ISM. We propose that the \wow was caused by the abrupt brightening of the hydrogen line of these clouds triggered by an intense radiation source, such as a magnetar flare or a soft gamma repeater (SGR).

The brightening of the cloud is similar to that from a maser flare but short-lived, from seconds to minutes. Astrophysical superradiance and masers involve coherent radiation, but they differ fundamentally in their mechanisms. Superradiance tends to produce short, intense bursts, whereas masers are associated with continuous or pulsed emission and are more commonly observed in various astrophysical environments. In particular, we showed that Dicke's supperradiance can explain the intensity and duration of the signal. The \wow might signify the first documented instance of an astrophysical maser-like flare or supperradiance of the 21-cm atomic hydrogen line.

Our hypothesis explains all the discernible attributes of the \wow, namely its narrowband frequency, duration, intensity, absence of modulation, and rarity. However, it does not conclusively rule out the possibility that the signal may have originated from alternative sources, encompassing both artificial terrestrial or extraterrestrial origins.

We advocate for a thorough examination of archival data as well as new observational efforts to accurately pinpoint the provenance of the signal or identify analogous \textit{wow signals} events. Since these \HI clouds are easy to recognize, it may be possible to determine the exact location of the source of the \wow. Identifying the trigger source, however, would prove challenging. It may be close or behind the cloud location, or far in the background.

These signals are likely to escape detection, not only due to their scarcity and association with small \HI clouds, but also because contemporary observational strategies typically focus on broadband signals, prolonged integration times, or target much narrower signal bandwidths. It was not until we engaged in the search for 10 kHz narrowband signals, utilizing a drift (or mapping) mode, that the small \HI clouds came to our attention as a possible source for the \wow.

Our study did not conclude that the \wow constituted evidence of a signal emanating from an extraterrestrial civilization. However, null results are instrumental in refining future technosignature searches. NASA and other agencies remain committed to their search for biosignatures within the atmospheres of potentially habitable exoplanets \citep{2018AsBio..18..663S}.

The next question after a positive biosignature detection would be whether it is also associated with intelligent life. It would be unsatisfactory to claim ignorance due to the absence of interest on technosignature searches. Therefore, we also emphasize the need for more sustained, long-term technosignature searches, particularly near the hydrogen line. Prospective extraterrestrial civilizations may similarly acknowledge the scarcity of natural \HI masers and could potentially exploit them to garner attention. However, it is plausible that they also deem it rude to direct a maser toward another entity. 

Our findings highlight once again the importance of reanalyzing historical datasets with different or more advanced methods, which could potentially uncover subtleties missed in previous analyses. For example, fast radio bursts (FRBs) were discovered in archival data and later confirmed by observations \citep{2007Sci...318..777L}. In our study, the replication of the observation and analysis methods used by Ohio SETI was the key to the association with \HI clouds.

Our Arecibo Wow! project will continue examining the rest of our datasets, which consist of numerous hours of observing stars with planets at frequencies ranging from 1 to 10 GHz. We will expand our narrowband search to incorporate artificially dispersed and modulated wideband signals, among other techniques \citep{2020RAA....20...78L}. Subsequently, we will try to move the search to over fifty decades of data (3 petabytes) from the Arecibo Observatory, now archived and available to the scientific community at the \href{https://tacc.utexas.edu/}{Texas Advanced Computing Center} (TACC). We also anticipate the dissemination of calibrated and reduced datasets of Arecibo Wow! to the broader scientific community.

The Arecibo Observatory is currently a center for education and research in science, technology, engineering, and mathematics (STEM) known as the \href{https://www.areciboc3.org/}{Arecibo C3}. Looking to the far future, the facility may potentially resume its role in astronomical observations through initiatives such as the Next Generation Arecibo Telescope (nGAT) or participation as a site for the Next Generation Very Large Array (ngVLA) \citep{2021arXiv210301367A,2024arXiv240814497W}. In the near term, a reestablishment of its 2.5 to 14 GHz 12-meter telescope should provide new opportunities for astronomical research and education \citep{2024RaSc...5907839R}.

\begin{acknowledgments}
\section*{Acknowledgments}
We extend our gratitude to former staff members of the Arecibo Observatory, Tapasi Ghosh and Christopher Salter, for helping us with our observation protocols and for their valuable comments and suggestions. We are grateful to Anish Roshi Damodaran (UCF) for his assistance with data calibration and comments. We thank Bob Dixon (formerly Big Ear staff), Robert F. Minchin (NRAO), Jean-Luc Margot (UCLA), Joshua Peek (APL), and Joel Weisberg (Carleton College) for valuable comments on the manuscript.

We also thank the computational resources provided by the High Performance Computing Facility (HPCf) of the University of Puerto Rico (UPR), Juan Padilla (SoftwareOne), Amazon Web Services (AWS), and all the people who contributed to save our Arecibo Observatory data. Finally, we thank Edgard Rivera-Valentín (JHU-APL) for encouraging us to use the Arecibo Observatory because it was easy to use; not true, but too late for that.
\end{acknowledgments}

%\vspace{5mm}

\facility{Arecibo Observatory, Big Ear}

\software{\href{https://www.naic.edu/arecibo/}{The Arecibo Observatory Software Library}, \href{https://github.com/wlandsman/IDLAstro}{IDLAstro}, \href{https://www.astropy.org/}{Astropy} \citep{astropy:2013, astropy:2018, astropy:2022}, \href{https://aladin.cds.unistra.fr/}{Aladin Sky Atlas}.}

\bibliography{awowi-v1}{}
\bibliographystyle{aasjournal}

\end{document}

\title{\large{Supplementary Tables and Figures} \\ \small{Arecibo Wow! I: An Astrophysical Explanation for the \wow}}

\author[0000-0002-0786-7307]{Abel M\'endez}
\affiliation{Planetary Habitability Laboratory, University of Puerto Rico at Arecibo}%, PO Box 4010, Arecibo, PR 00613, USA}

\correspondingauthor{Abel M\'endez}
\email{abel.mendez@upr.edu}

\author[0000-0003-3455-8814]{Kevin Ortiz Ceballos}
\affiliation{Center for Astrophysics ${\rm \mid}$ Harvard {\rm \&} Smithsonian}%, 60 Garden St, Cambridge, MA 02138, USA}

\author[0000-0002-6140-3116]{Jorge I. Zuluaga}
\affiliation{SEAP/FACom, Instituto de F\'{\i}sica - FCEN, University of Antioquia}%Calle 70 No. 52-21, Medell\'in, Colombia

\section{Introduction}

This material provides additional tables or figures to enhance and support the Arecibo Wow! observations.

%%%%%%%%%%%%%%%%%%%%%%%%%%%%%%%%%%%%%%%%%%%%%%%%%%%%%%%%%%%%%%%%%%%%%%%%%%%%%%%%%
%   Targets Table
%%%%%%%%%%%%%%%%%%%%%%%%%%%%%%%%%%%%%%%%%%%%%%%%%%%%%%%%%%%%%%%%%%%%%%%%%%%%%%%%%

\begin{deluxetable}{llrrlcc}
\tablewidth{0pt}
\tablecaption{Observation dates and scans of targets in drift mode and in the L-band. These and other targets were also observed in other modes, including calibrators. The date (AST) and scan ID columns are provided to facilitate the identification of the Arecibo FITS files with these data sets. These observations are part of the Arecibo Observatory project A3123.\label{tab:suppl-scans}}
\tablehead{
\colhead{Date (AST)} & \colhead{Date (MJD)} & \colhead{RA (deg)} & \colhead{DEC (deg)} & \colhead{Target} & \colhead{Scan ID} & \colhead{Time (min)}
}
\startdata
20200210  & 58889.960405  & 43.1284   & 16.8811  & Teegarden   & 52 & 1 \\
\hline
20200225  & 58904.852072  & 42.7526   & 16.8806  & Teegarden   & 18 & 4 \\
          & 58904.886898  & 42.0009   & 16.8797  & Teegarden   & 33 & 10 \\
          & 58904.895660  & 45.1845   & 16.4455  & B0300+162   & 36 & 5 \\
          & 58904.901667  & 42.6273   & 16.8805  & Teegarden   & 39 & 5 \\
          & 58904.914306  & 44.3753   & 16.9992  & Void        & 52 & 5 \\
          & 58904.920046  & 42.6290   & 16.8805  & Teegarden   & 55 & 5 \\
\hline
20200430  & 58969.297257  & 269.3267  & 4.6936   & Barnard     & 18 & 1 \\
          & 58969.298576  & 269.3267  & 4.6936   & Barnard     & 21 & 1 \\
          & 58969.299896  & 269.3267  & 4.6936   & Barnard     & 24 & 1 \\
          & 58969.361435  & 293.6067  & 21.8969  & SGR1935+2154& 62 & 1 \\
          & 58969.362789  & 293.6067  & 21.8969  & SGR1935+2154& 65 & 1 \\
\hline
20200506  & 58975.263692  & 269.5042  & 13.4759  & B1756+134   & 6  & 1 \\
          & 58975.270880  & 261.3131  & 2.1117   & HD157881    & 13 & 1 \\
          & 58975.279387  & 269.3267  & 4.6936   & Barnard     & 24 & 1 \\
          & 58975.344491  & 293.6067  & 21.8969  & SGR1935+2154& 60 & 1 \\
          & 58975.349734  & 293.6067  & 21.9219  & SGR1935+2154& 67 & 1 \\
          & 58975.351389  & 293.6067  & 21.8969  & SGR1935+2154& 70 & 1 \\
          & 58975.352697  & 293.6067  & 21.9219  & SGR1935+2154& 73 & 1 \\
          & 58975.354005  & 293.6068  & 21.9469  & SGR1935+2154& 76 & 1 \\
\hline
20200513  & 58982.307755  & 269.3268  & 4.5769   & Barnard     & 37 & 1 \\
          & 58982.309086  & 269.3269  & 4.6061   & Barnard     & 40 & 1 \\
          & 58982.310417  & 269.3270  & 4.6353   & Barnard     & 43 & 1 \\
          & 58982.311748  & 269.3270  & 4.6644   & Barnard     & 46 & 1 \\
          & 58982.313067  & 269.3271  & 4.6936   & Barnard     & 49 & 1 \\
          & 58982.314398  & 269.3271  & 4.7228   & Barnard     & 52 & 1 \\
          & 58982.315718  & 269.3272  & 4.7519   & Barnard     & 55 & 1 \\
          & 58982.317037  & 269.3272  & 4.7811   & Barnard     & 58 & 1 \\
          & 58982.322338  & 293.6068  & 21.7803  & SGR1935+2154& 61 & 1 \\
          & 58982.323646  & 293.6068  & 21.8094  & SGR1935+2154& 64 & 1 \\
          & 58982.324954  & 293.6069  & 21.8386  & SGR1935+2154& 67 & 1 \\
          & 58982.326262  & 293.6070  & 21.8678  & SGR1935+2154& 70 & 1 \\
          & 58982.327569  & 293.6070  & 21.8969  & SGR1935+2154& 73 & 1 \\
          & 58982.328877  & 293.6071  & 21.9261  & SGR1935+2154& 76 & 1 \\
          & 58982.330185  & 293.6071  & 21.9553  & SGR1935+2154& 79 & 1 \\
          & 58982.331493  & 293.6072  & 21.9844  & SGR1935+2154& 82 & 1 \\
\enddata
\end{deluxetable}

%%%%%%%%%%%%%%%%%%%%%%%%%%%%%%%%%%%%%%%%%%%%%%%%%%%%%%%%%%%%%%%%%%%%%%%%%%%%%%%%%
%   HI4PI Plots
%%%%%%%%%%%%%%%%%%%%%%%%%%%%%%%%%%%%%%%%%%%%%%%%%%%%%%%%%%%%%%%%%%%%%%%%%%%%%%%%%

\begin{figure}
    \centering
    \subfigure[Teegarden's Star.]{
        \includegraphics[width=0.23\textwidth]{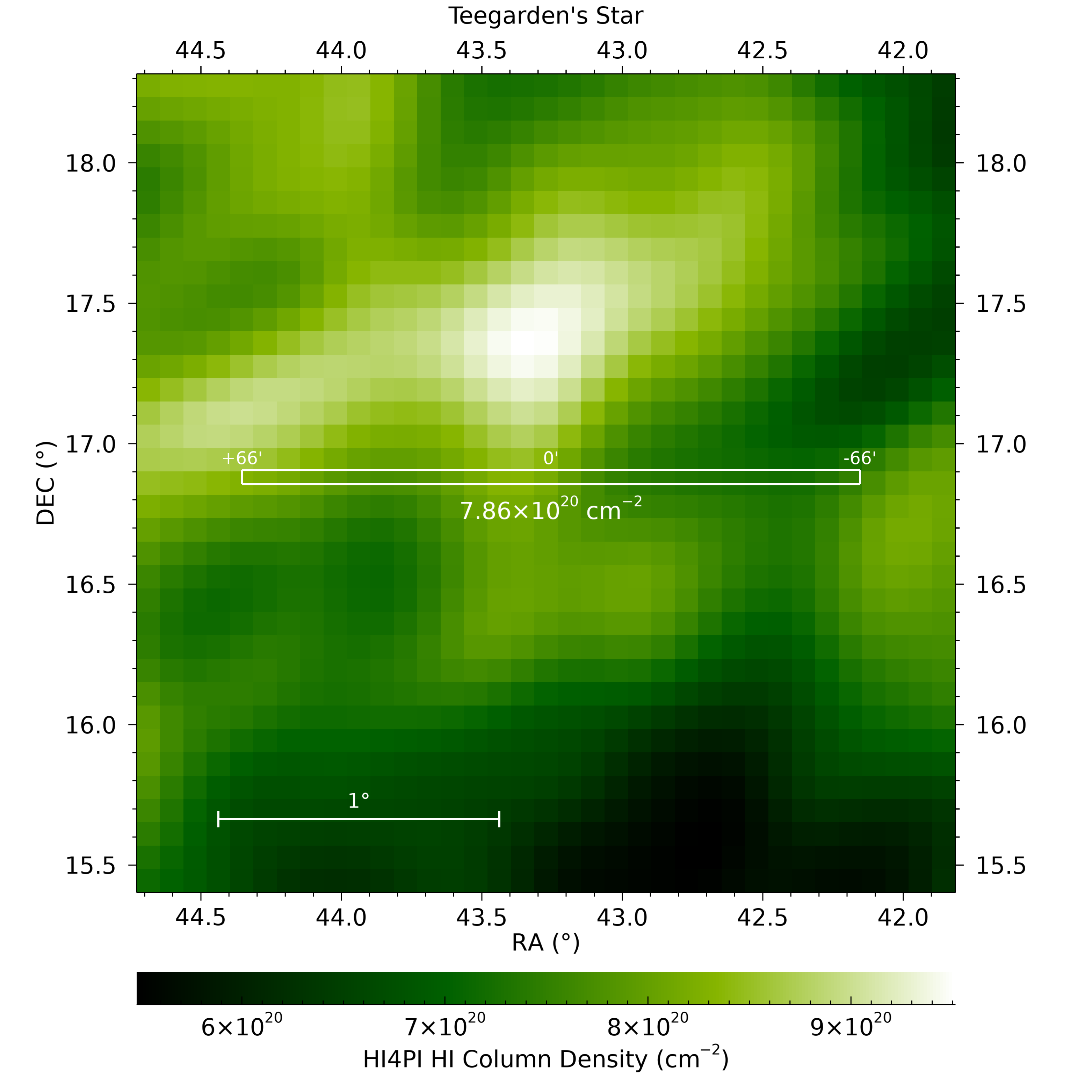}
    }
    \subfigure[Barnard's Star.]{
        \includegraphics[width=0.23\textwidth]{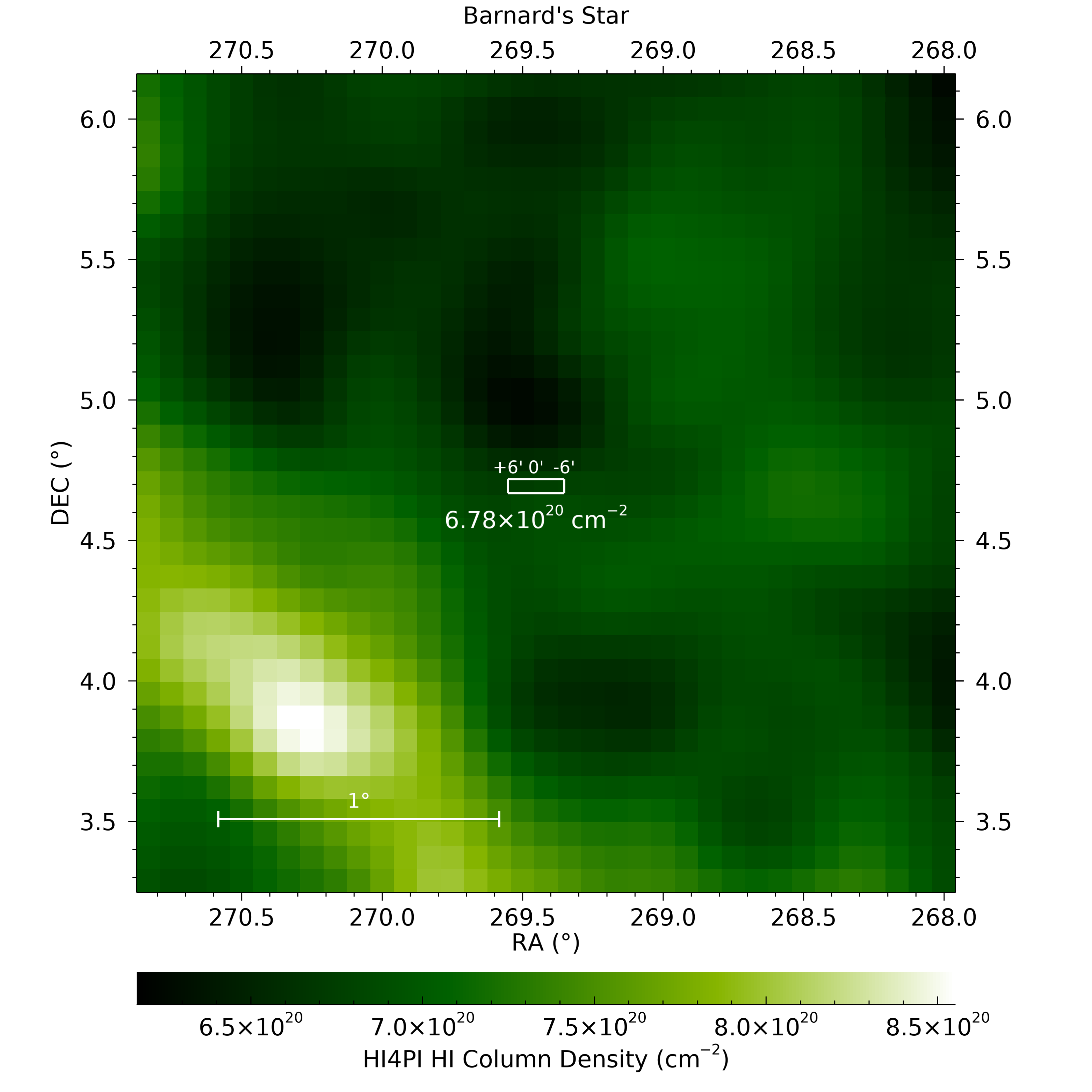}
    }
    \subfigure[HD 157881.]{
        \includegraphics[width=0.23\textwidth]{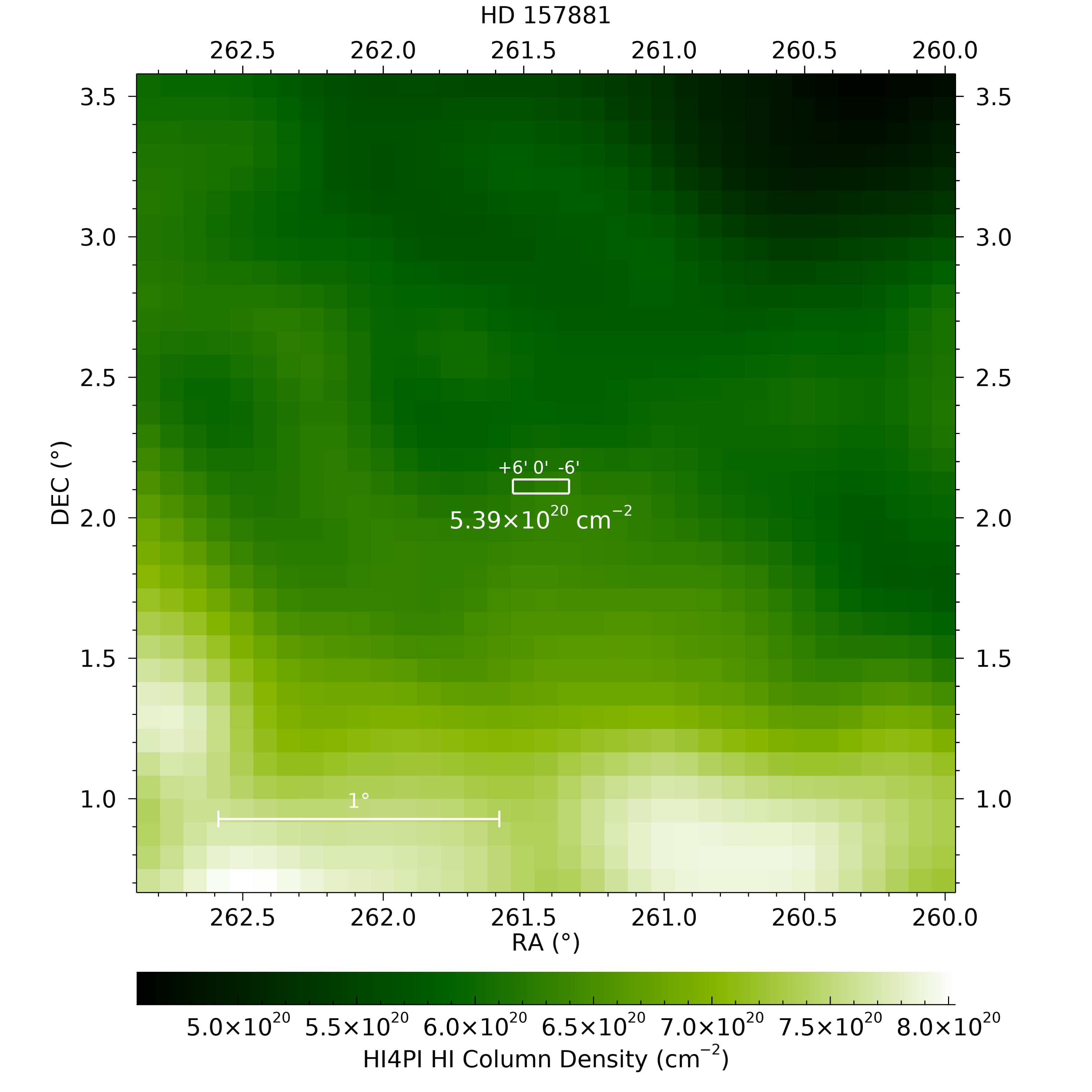}
    }
    \subfigure[SGR 1935+2154.]{
        \includegraphics[width=0.23\textwidth]{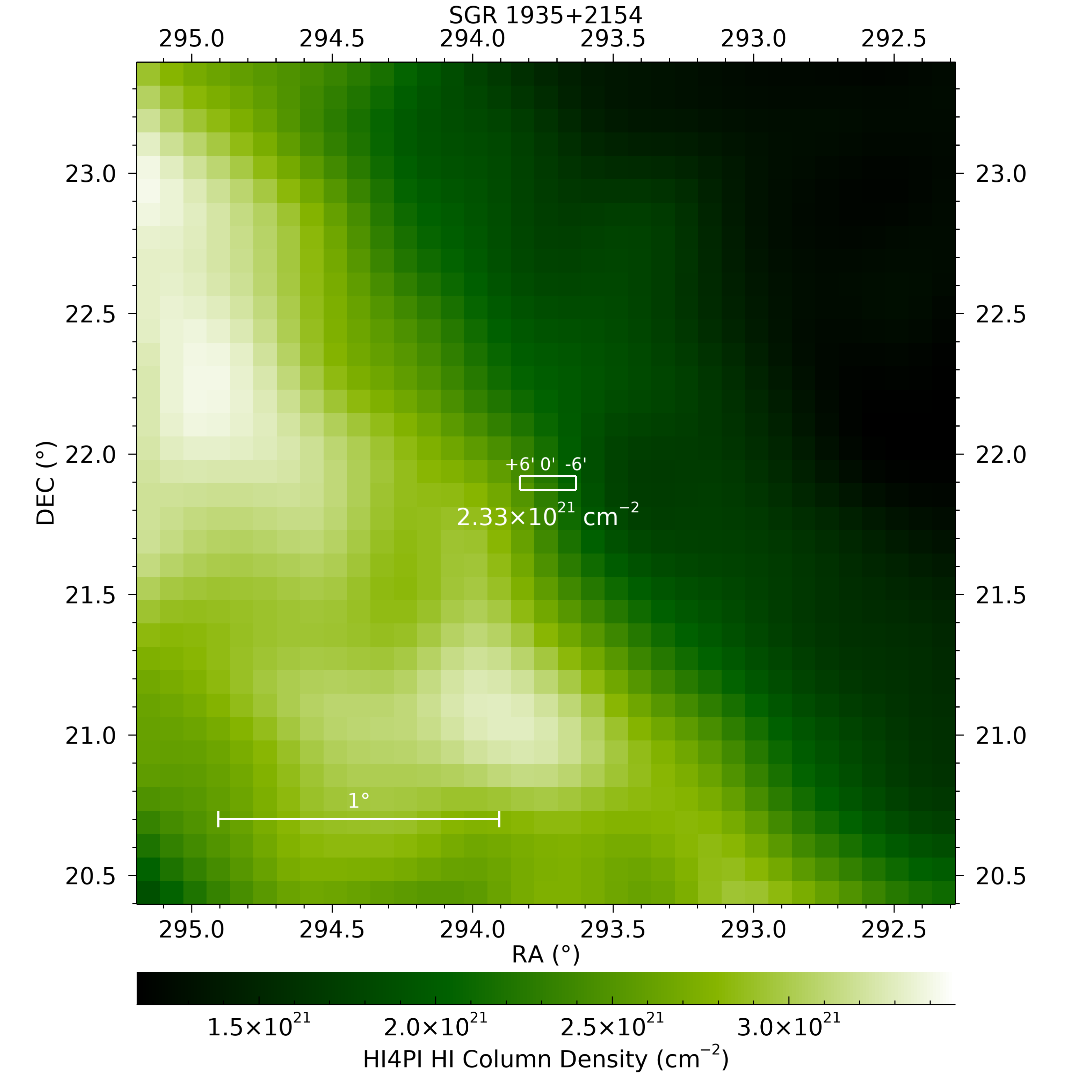}
    }
    \subfigure[3C 76.1 (B0300+162).]{
        \includegraphics[width=0.23\textwidth]{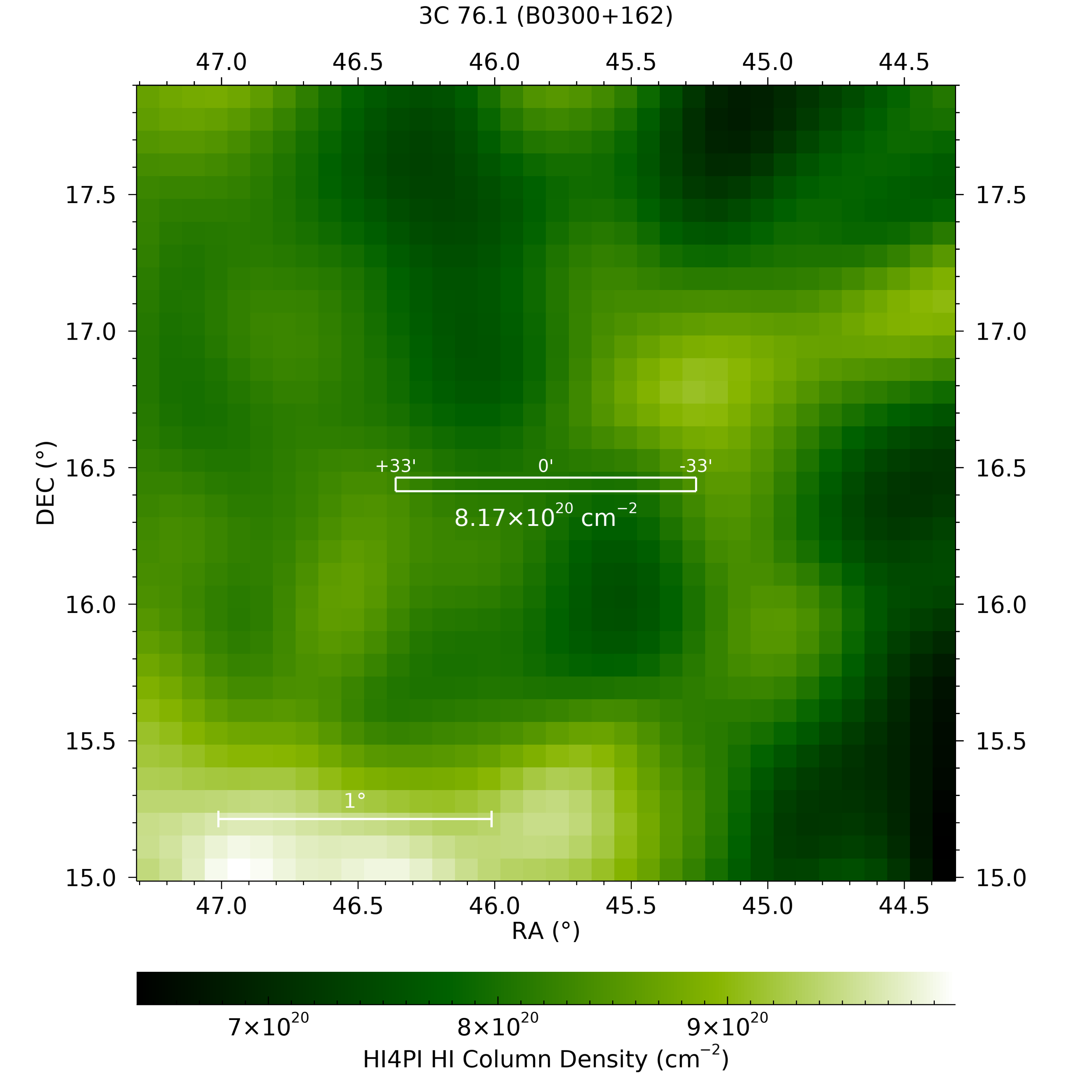}
    }
    \subfigure[3C 365 (B1756+134).]{
        \includegraphics[width=0.23\textwidth]{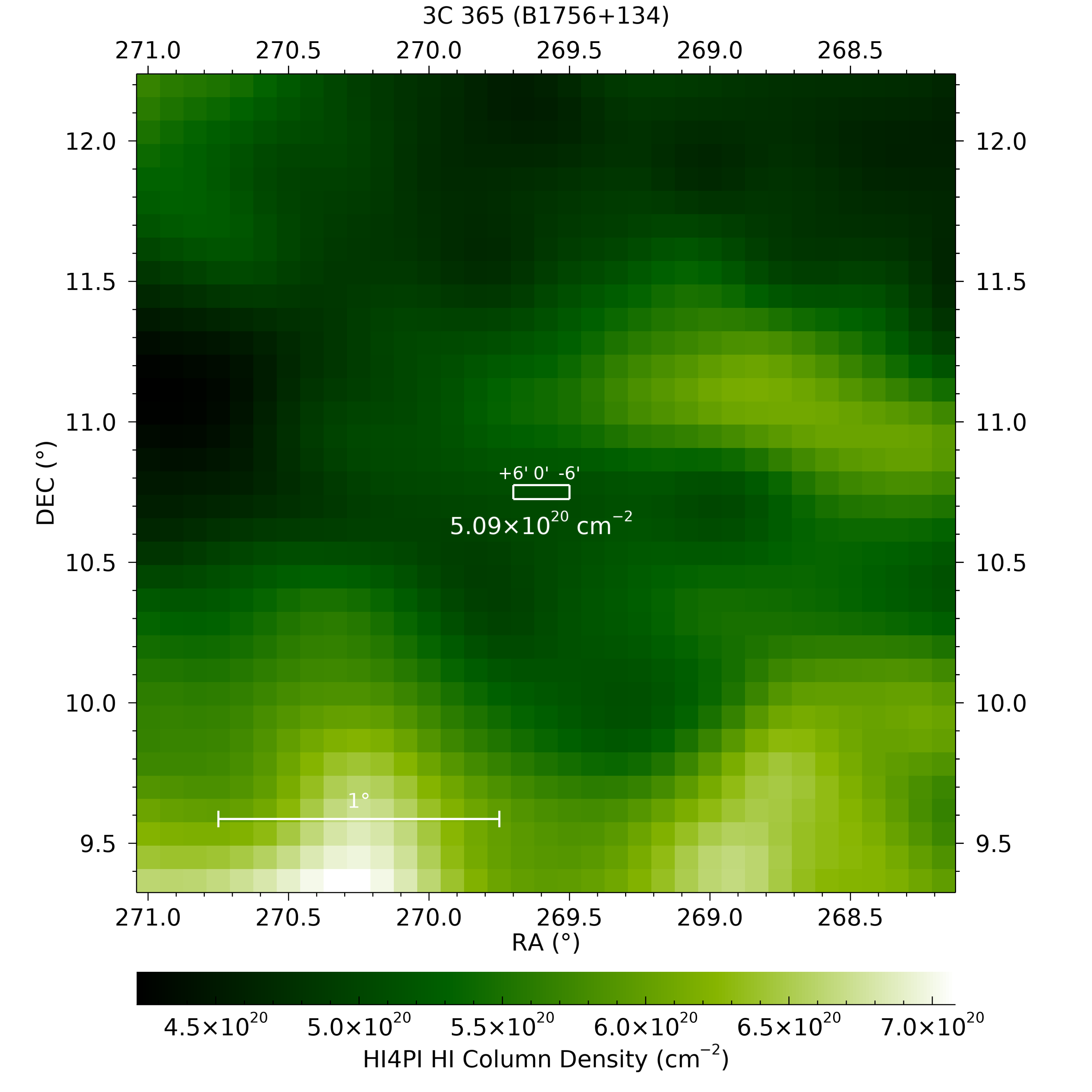}
    }
    \subfigure[Void.]{
        \includegraphics[width=0.23\textwidth]{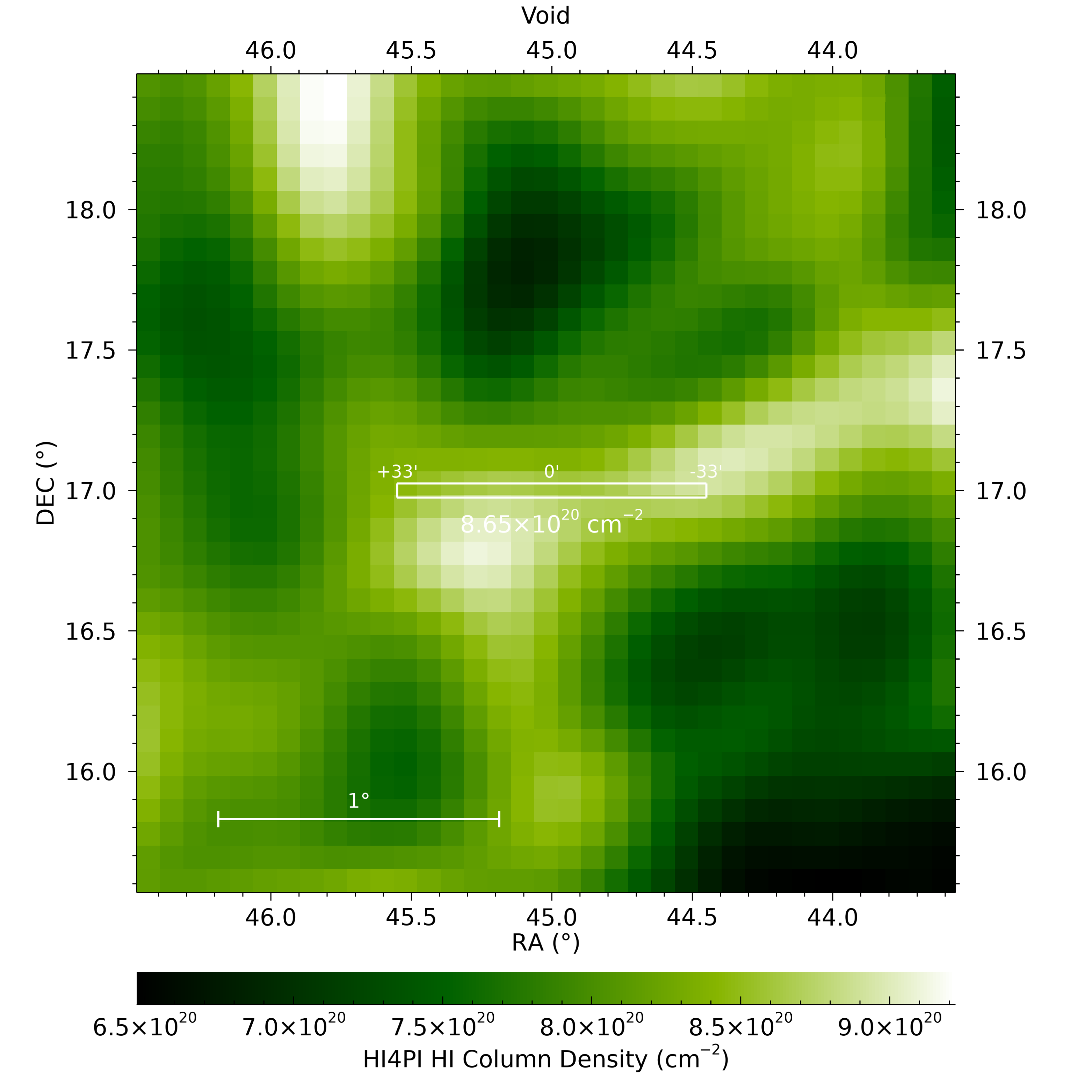}
    }
    \subfigure[\wow.]{
        \includegraphics[width=0.23\textwidth]{images/HI4PI/HI4PI_nhi_Wow_Signal.png}
    }
    \caption{HI4PI hydrogen column density between -10 and +10 km/s near all observed fields (figures a to g) compared to the \wow field (figure h). The scan strip of our observations and the average density are shown as in the middle of the plot (white box and text). The \wow figure shows the beamsize of the two horns used by the Big Ear. It is not clear in which of these horns the signal was detected. Note that the colorbar is not on the same scale in each plot to highlight the relative column density. Data from the \cite{2016A&A...594A.116H}.\label{fig:suppl-4HPI}}
\end{figure}

%%%%%%%%%%%%%%%%%%%%%%%%%%%%%%%%%%%%%%%%%%%%%%%%%%%%%%%%%%%%%%%%%%%%%%%%%%%%%%%%%
%   GALFA Plots
%%%%%%%%%%%%%%%%%%%%%%%%%%%%%%%%%%%%%%%%%%%%%%%%%%%%%%%%%%%%%%%%%%%%%%%%%%%%%%%%%

\begin{figure}
    \centering
    \subfigure[Teegarden's Star.]{
        \includegraphics[width=0.23\textwidth]{images/GALFA/GALFA_nhi_Teegardens_Star.png}
    }
    \subfigure[Barnard's Star.]{
        \includegraphics[width=0.23\textwidth]{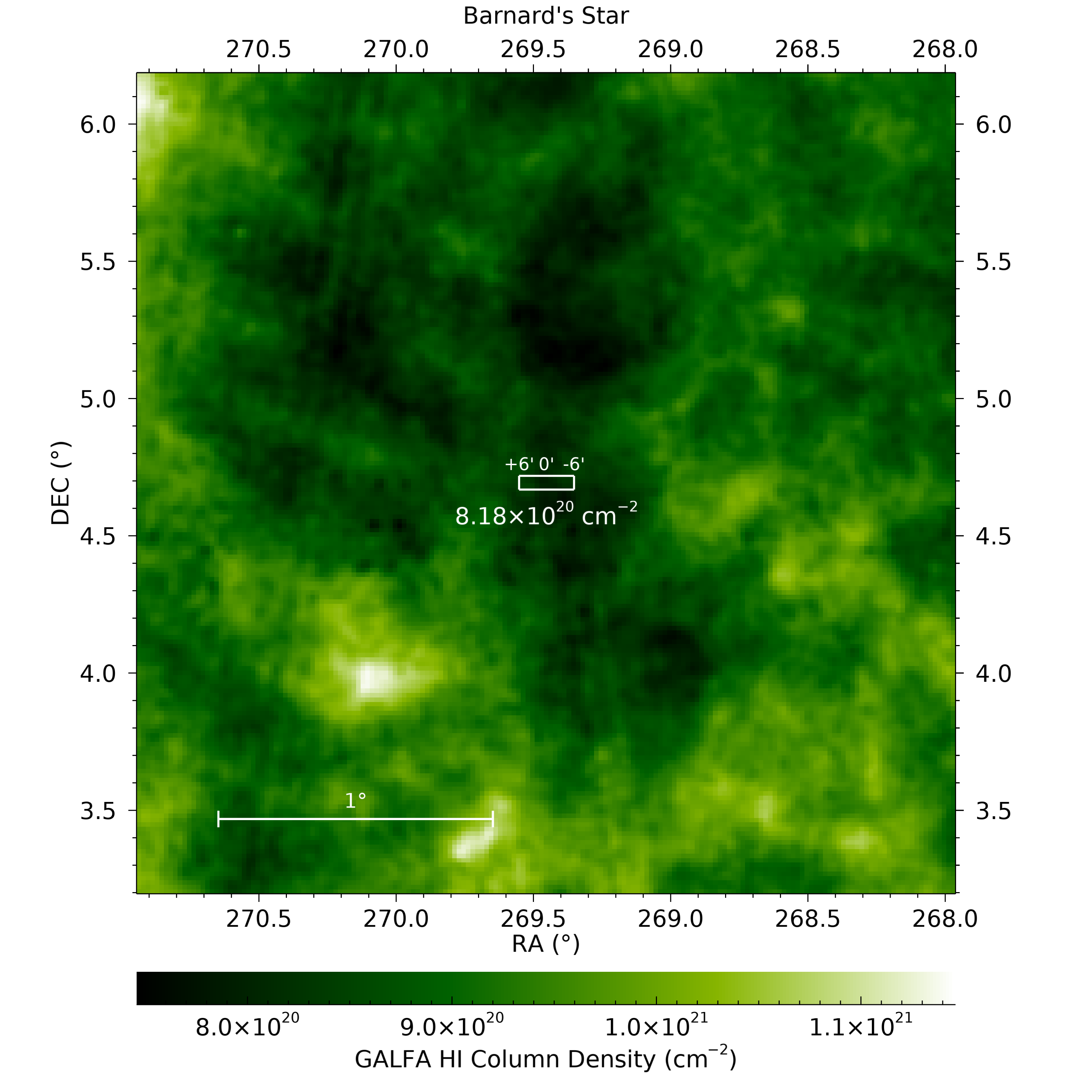}
    }
    \subfigure[HD 157881.]{
        \includegraphics[width=0.23\textwidth]{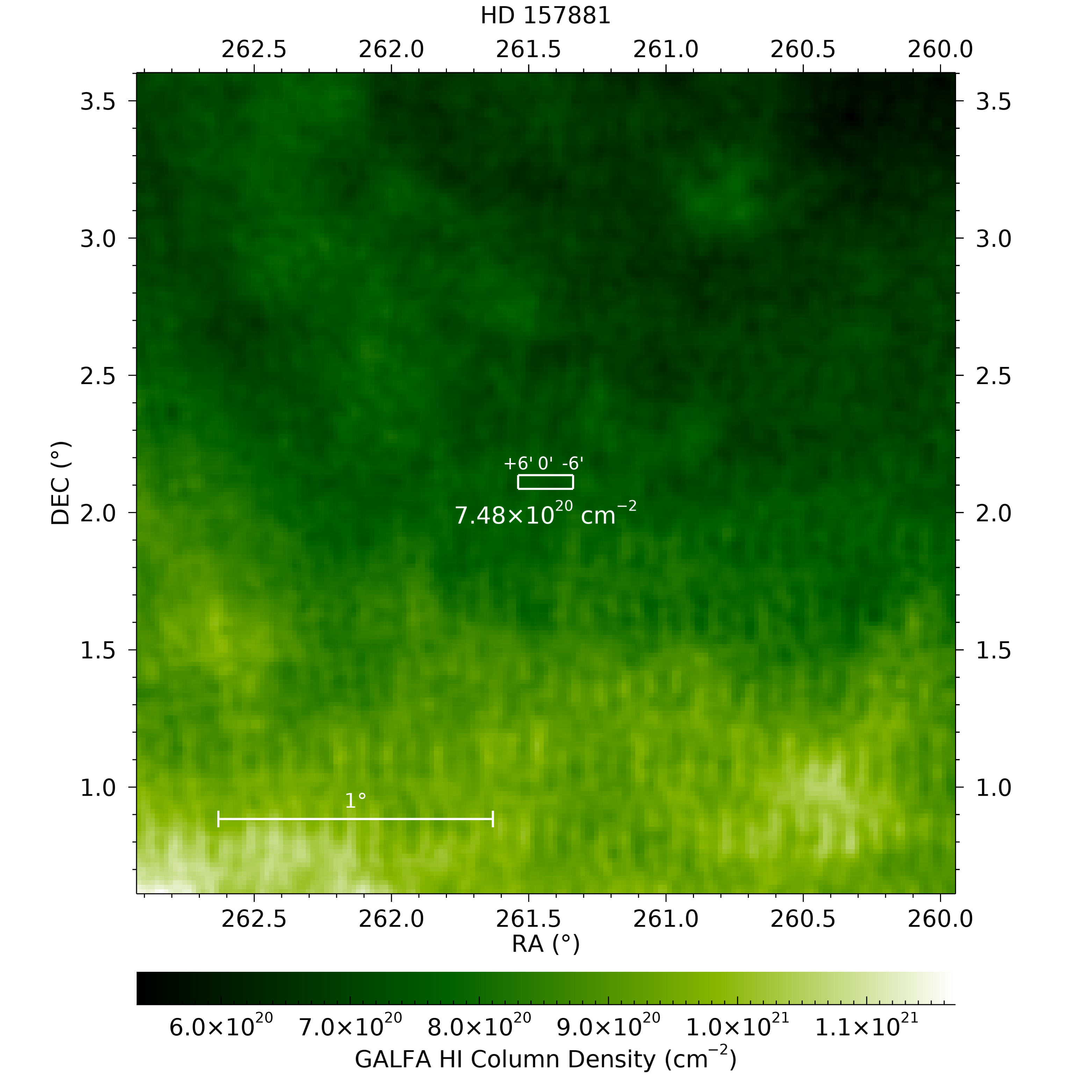}
    }
    \subfigure[SGR 1935+2154.]{
        \includegraphics[width=0.23\textwidth]{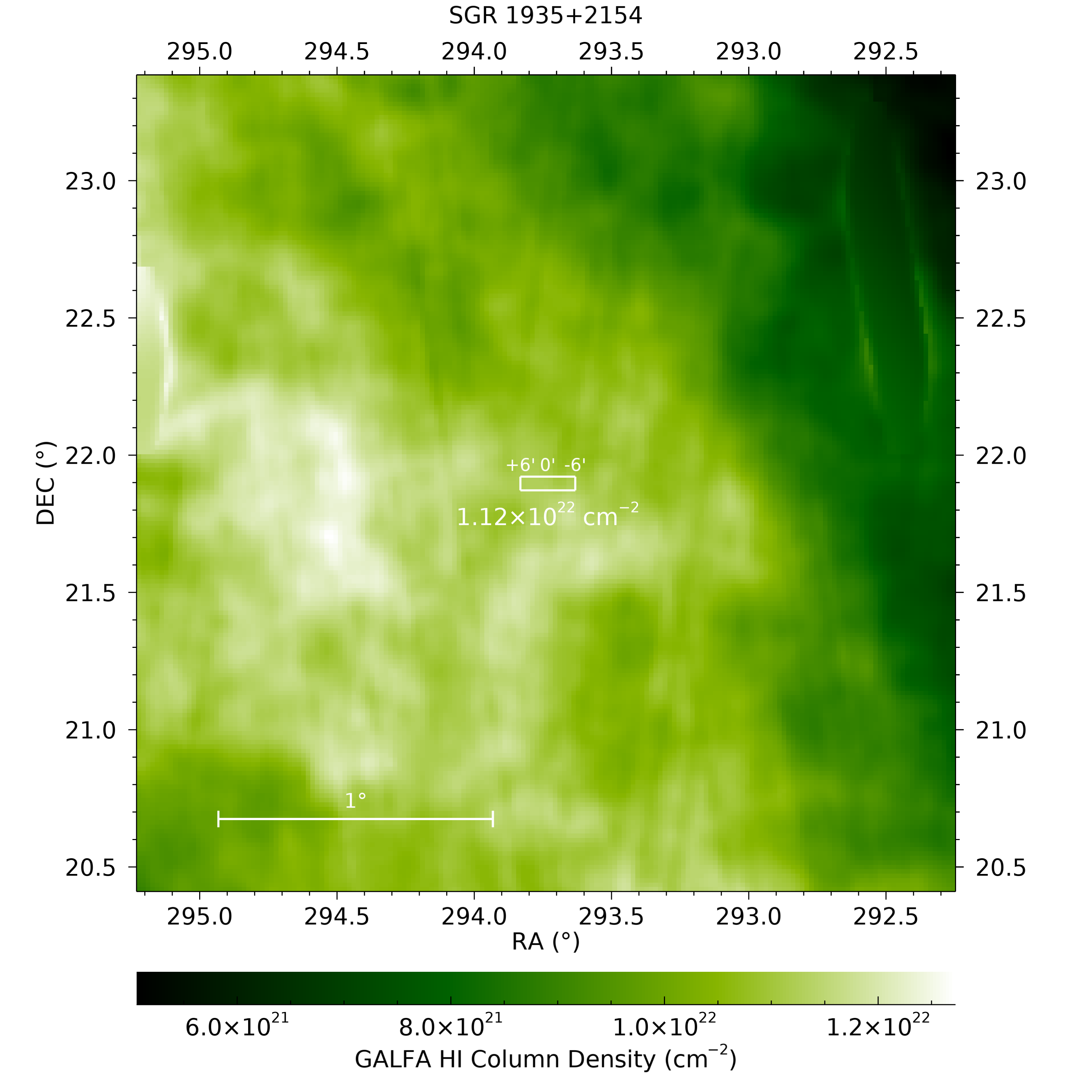}
    }
    \subfigure[3C 76.1 (B0300+162).]{
        \includegraphics[width=0.23\textwidth]{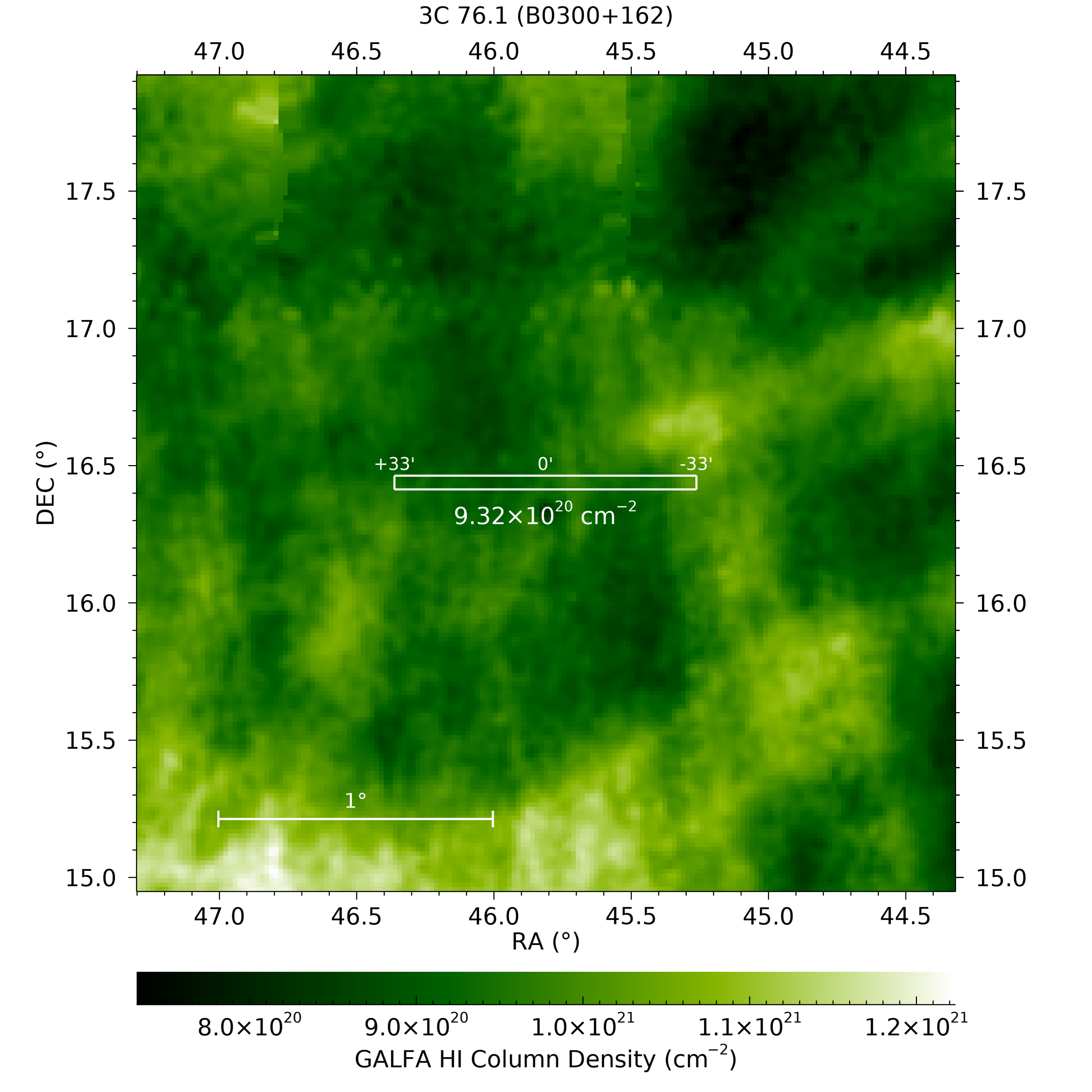}
    }
    \subfigure[3C 365 (B1756+134).]{
        \includegraphics[width=0.23\textwidth]{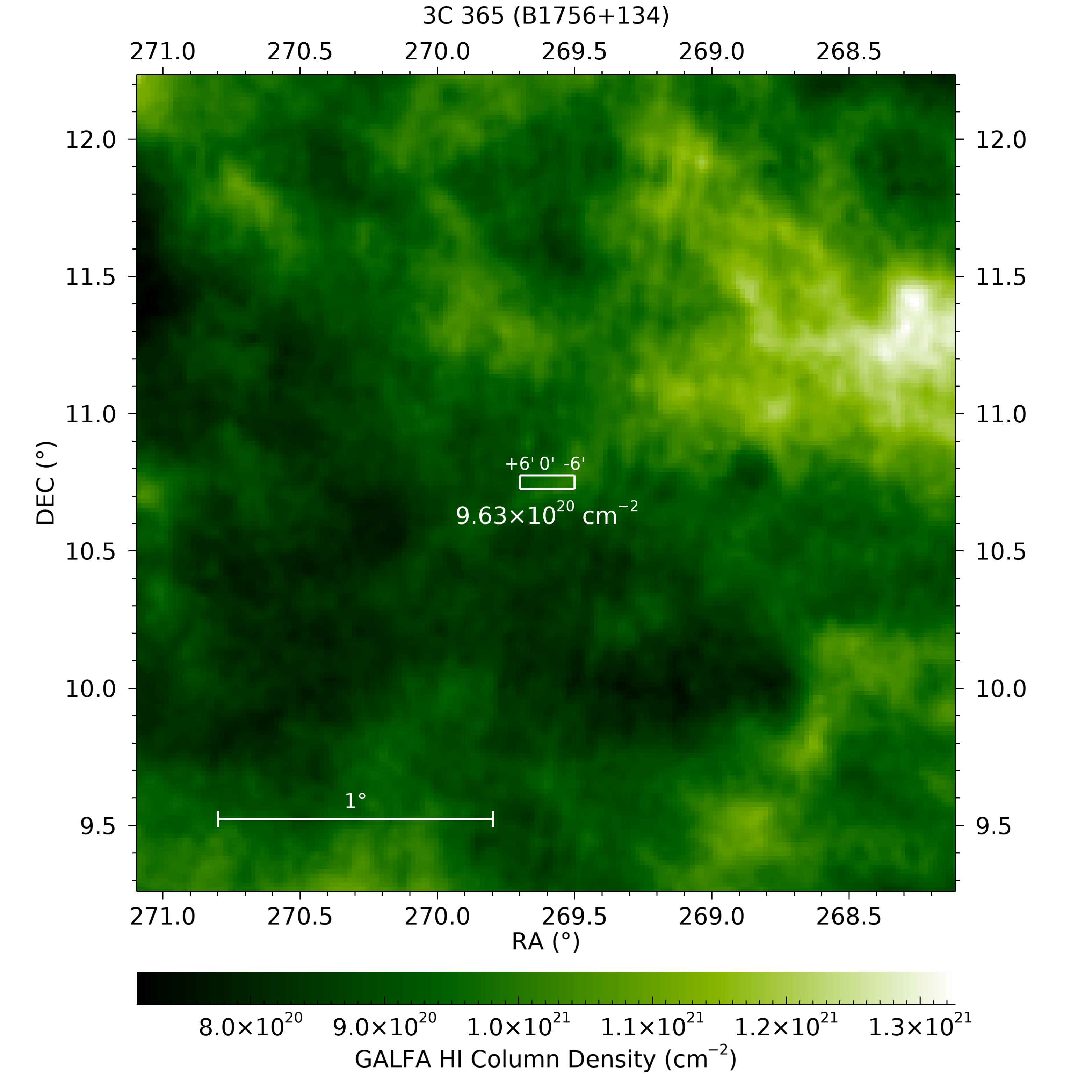}
    }
    \subfigure[Void.]{
        \includegraphics[width=0.23\textwidth]{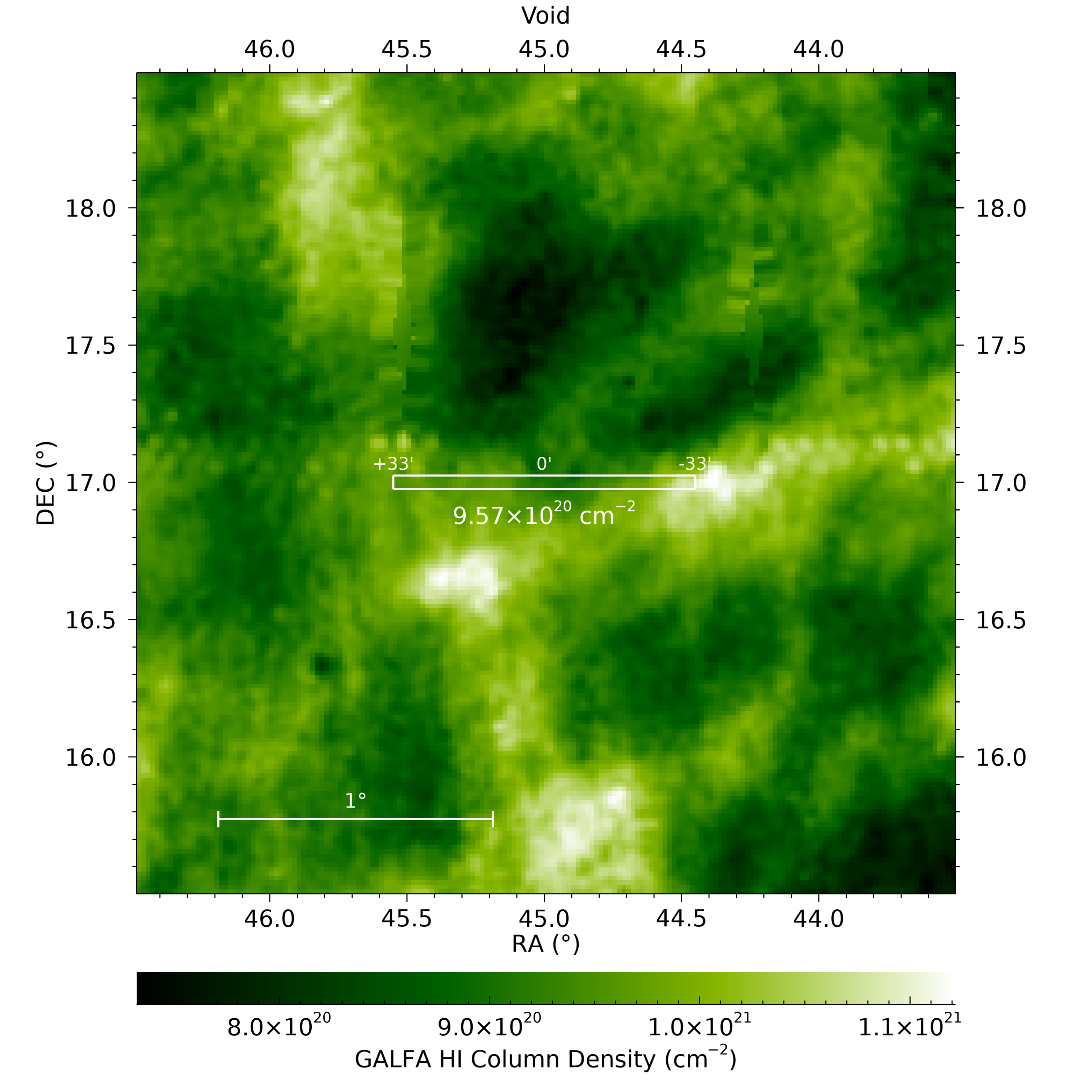}
    }
    \caption{GALFA hydrogen column density between -10 and +10 km/s near all observed fields (figures a to g). The scan strip of our observations and the average density are shown as in the middle of the plot (white box and text). Note that the colorbar is not on the same scale in each plot to highlight the relative column density. Data from \cite{2018ApJS..234....2P,DVN/T9CFT8_2017}.\label{fig:suppl-GALFA}}
\end{figure}

%%%%%%%%%%%%%%%%%%%%%%%%%%%%%%%%%%%%%%%%%%%%%%%%%%%%%%%%%%%%%%%%%%%%%%%%%%%%%%%%%
%   Clouds Profile Plots
%%%%%%%%%%%%%%%%%%%%%%%%%%%%%%%%%%%%%%%%%%%%%%%%%%%%%%%%%%%%%%%%%%%%%%%%%%%%%%%%%

\begin{figure}
    \centering
    \subfigure[Cloud A.]{
        \includegraphics[width=0.45\textwidth]{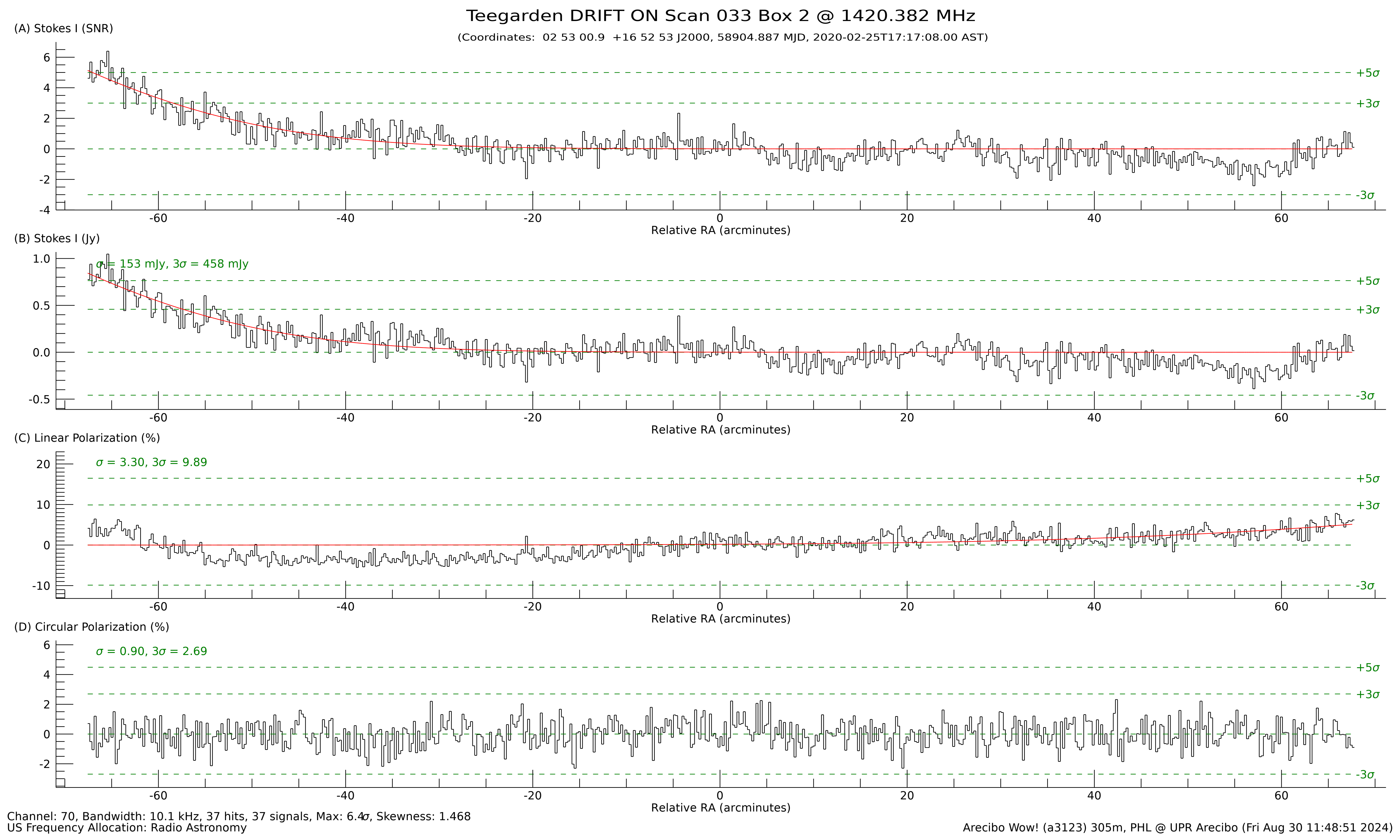}
    }
    \subfigure[Cloud B.]{
        \includegraphics[width=0.45\textwidth]{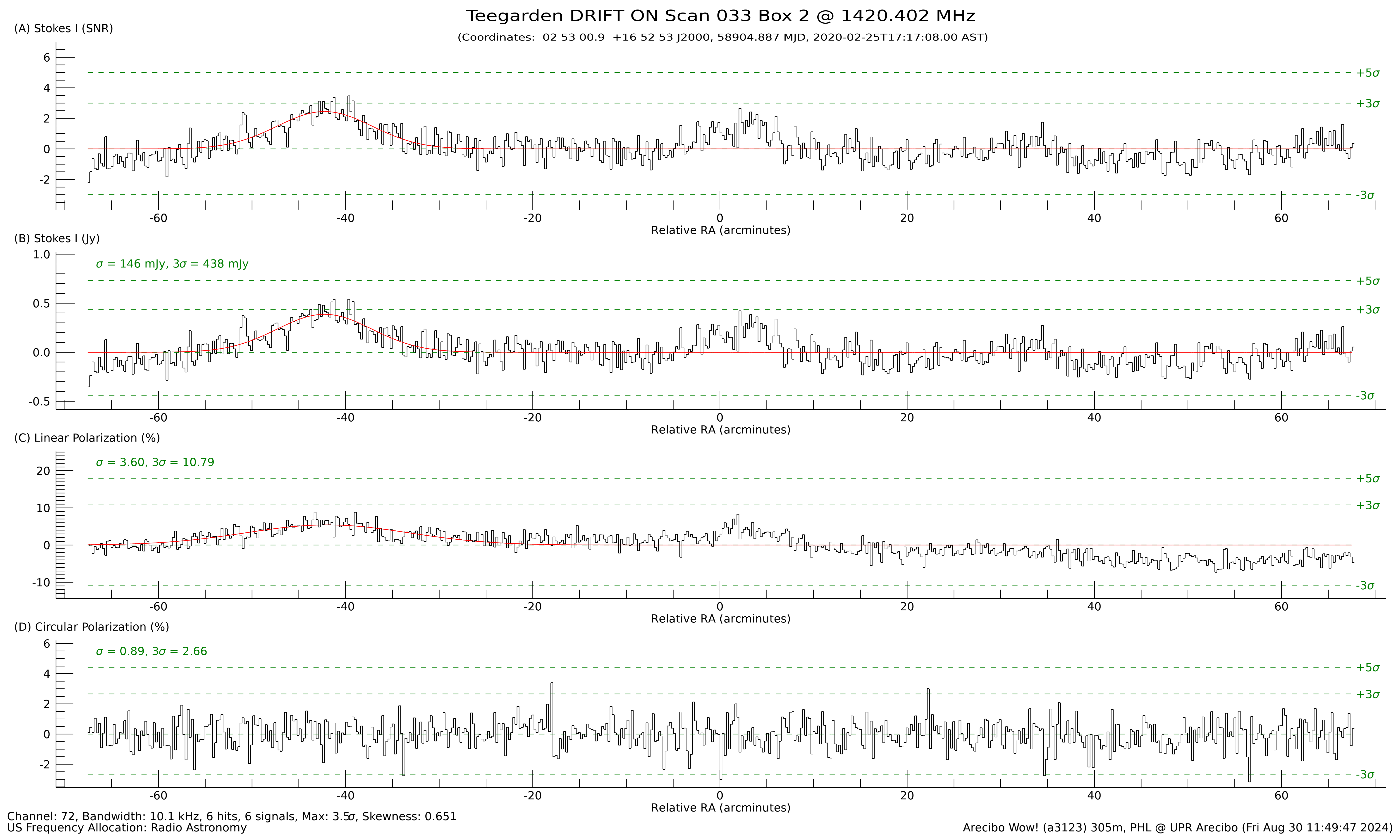}
    }
    \subfigure[Cloud C.]{
        \includegraphics[width=0.45\textwidth]{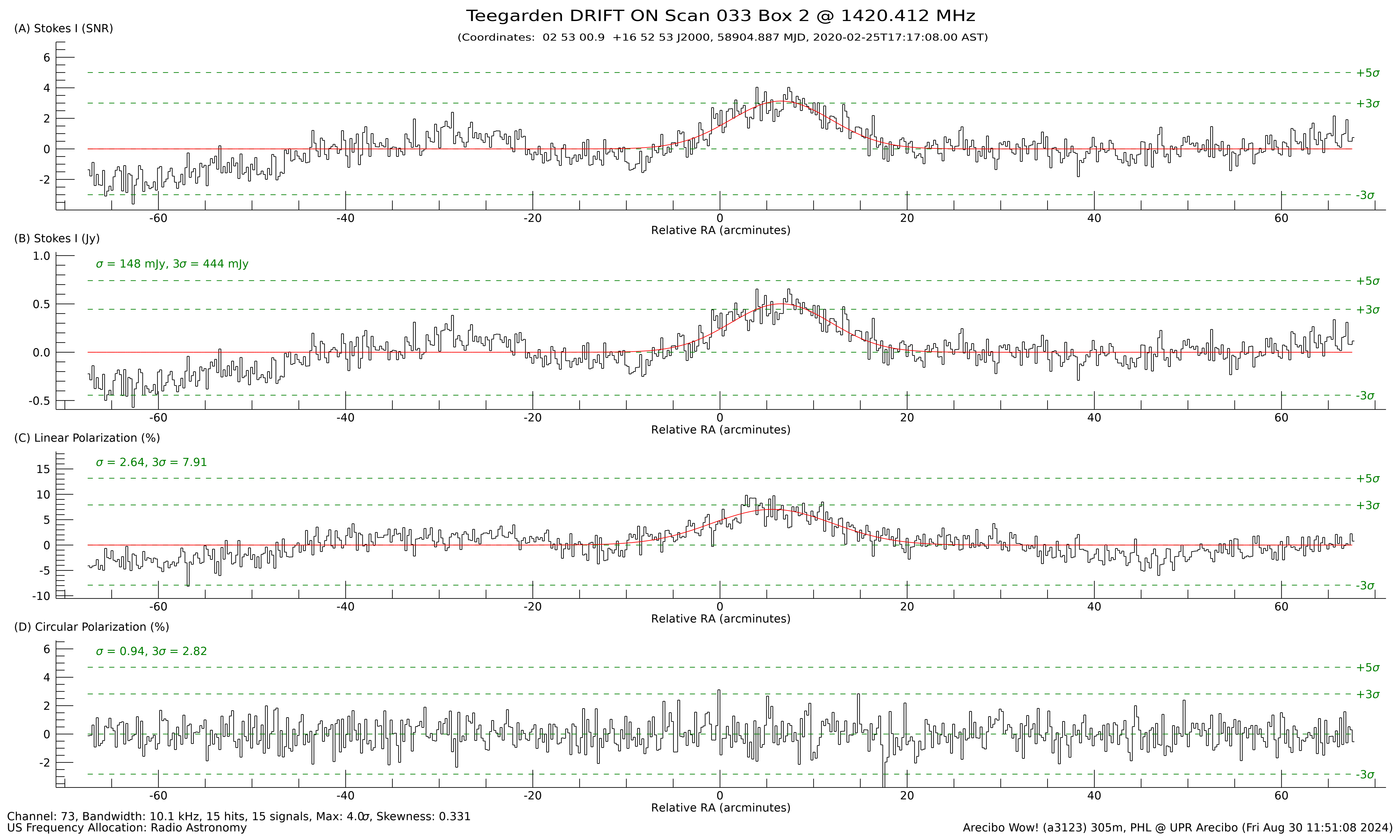}
    }
    \subfigure[Cloud D.]{
        \includegraphics[width=0.45\textwidth]{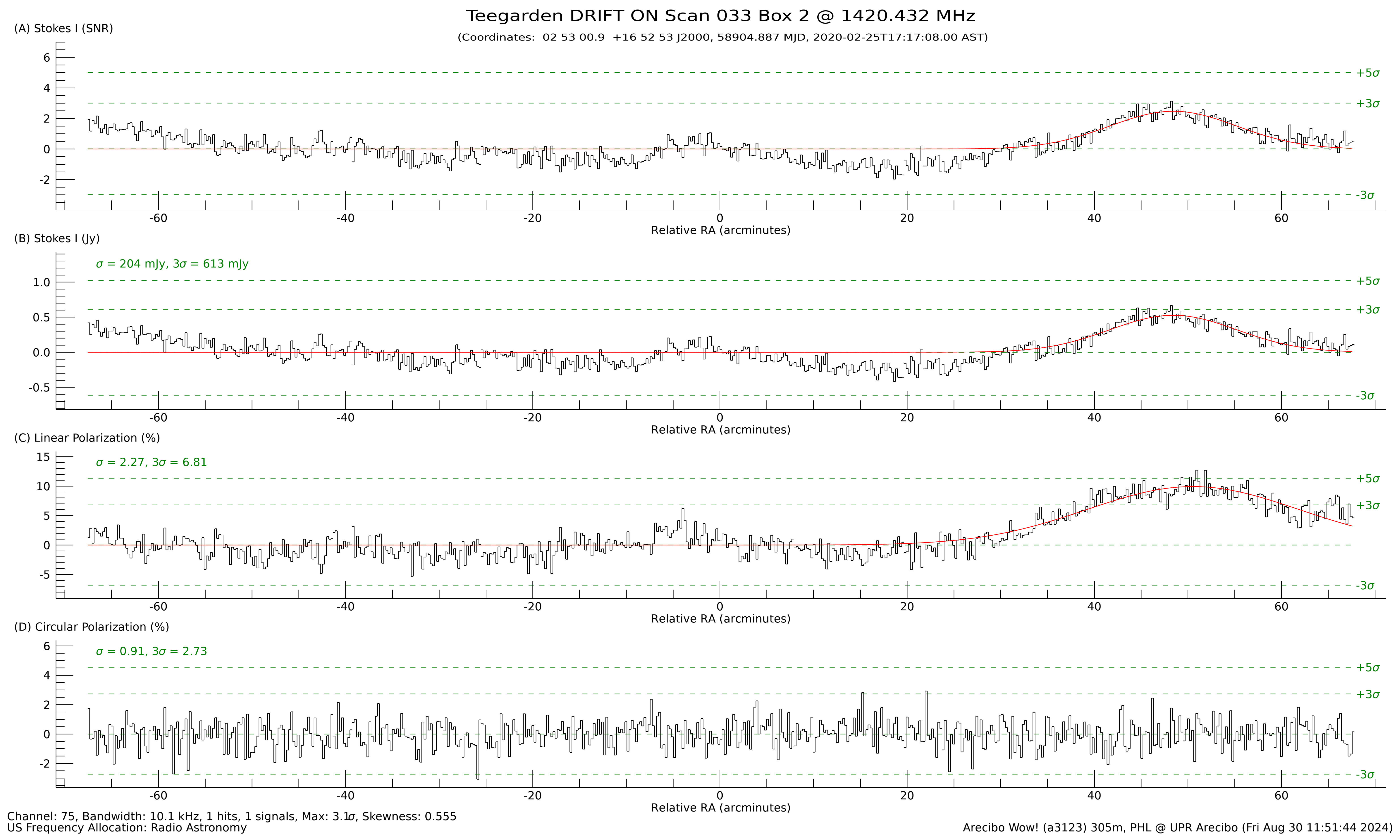}
    }
    \caption{Clouds profiles. Red lines show a fitted Gaussian function used to get their properties.}
    \label{fig:suppl-clouds}
\end{figure}

\bibliography{awowi-v1}{}
\bibliographystyle{aasjournal}